\documentclass[a4paper,11pt]{article}
\usepackage{jcappub} 
\usepackage{lineno}
\usepackage{orcidlink}

\usepackage[colorlinks=true,citecolor=blue,linkcolor=blue]{hyperref}
\usepackage{amsmath}
\usepackage{bm}
\usepackage{amssymb}
\usepackage{epsfig}
\usepackage{epstopdf}
\usepackage{url}
\usepackage{color}
\usepackage[dvipsnames]{xcolor}
\usepackage[utf8]{inputenc} 
\usepackage{amssymb}
\usepackage{slashed}
\usepackage{textcomp}
\usepackage{calc}
\usepackage{hyperref}
\usepackage{soul}
\usepackage{cancel}
\usepackage[normalem]{ulem}

\allowdisplaybreaks

\catcode`\@=11
\def\lsim{\mathrel{\mathpalette\@versim<}}
\def\gsim{\mathrel{\mathpalette\@versim>}}
\def\@versim#1#2{\vcenter{\offinterlineskip
\ialign{$\m@th#1\hfil##\hfil$\crcr#2\crcr\sim\crcr } }}
\catcode`\@=12

\DeclareMathOperator{\Tr}{Tr}
\newcommand{\al}[1]{\begin{align}#1\end{align}}

\newcommand{\be}{\begin{eqnarray}}
\newcommand{\ee}{\end{eqnarray}}

\makeatletter
\gdef\@fpheader{}
\makeatother

\begin{document}
\title{\boldmath\LARGE \vspace*{15mm}Gravity and the Hierarchy Problem}

\author[a,*]{\small Thede de Boer\,\orcidlink{0000-0002-9916-3306},}
\author[a,b]{Jisuke Kubo\,\orcidlink{0000-0003-2211-4685},}
\author[a]{Manfred Lindner\,\orcidlink{0000-0002-3704-6016},}
\author[a]{Markus Reinig\,\orcidlink{0009-0004-7339-2057}}
\affiliation[a]{Max-Planck-Institut f\"ur Kernphysik, Saupfercheckweg 1, 69117 Heidelberg, Germany}
\affiliation[b]{Department of Physics, University of Toyama, 3190 Gofuku, Toyama 930-8555, Japan\\}
%
\emailAdd{thede.deboer@mpi-hd.mpg.de}
\emailAdd{kubo@mpi-hd.mpg.de}
\emailAdd{lindner@mpi-hd.mpg.de}
\emailAdd{markus.reinig@mpi-hd.mpg.de}

\abstract {We propose a mechanism where the dynamical generation of the Planck mass in scale invariant gravity leads to Einstein gravity, successful inflation and an explanation of the hierarchy problem of the Standard Model. We will discuss the scale generation by dynamical symmetry breaking and phenomenological consequences.}

\notoc 
\flushbottom

\maketitle
%
\newpage 
\section{Introduction}
Both the Standard Model of particle physics (SM) and the standard cosmological model ($\Lambda$CDM) describe experimental observations in particle physics and cosmology extremely well~\cite{ParticleDataGroup:2024cfk}. For both of them exist experimental indications which point to new physics and both also have open conceptual questions or even theoretical problems. This led for both of them to many ideas for extensions and new physics which go beyond these standard models.

On the SM side exists among others the so-called hierarchy problem~\cite{Peskin:2025lsg} which is essentially the fact that a large separation of the masses of the SM Higgs field $H$ and of some new field required in extensions is unnatural within Quantum Field Theory (QFT). There are, however, important differences for scale invariant or conformal theories which we will discuss. If the SM is extended like in our model below in a scale invariant manner then there is also a hierarchy problem if a new fundamental scalar $\Phi$ is introduced. The main reason  is that a portal term $\lambda_pH^\dag H\Phi^\dag \Phi$ with the portal coupling $\lambda_p$ is not forbidden or protected by symmetry such that quantum effects are expected to push the mass $m_H$ of $H$ towards the much higher mass $M_\Phi$ of $\Phi$. A very important difference, which is essential for our paper, is that a tree level portal term does not exist if one of the scalars is effective (composite). We will see how a tiny effective portal term can then emerge if it is mediated only by gravitational interactions. 

Without a mechanism which explains and stabilizes a tiny portal the problem must be avoided in other ways. The problem can then, for example, be avoided by supersymmetry which postulates for each known field of the SM a partner with opposite statistics such that the quadratic interdependence of quantum effects is systematically cancelled. Supersymmetric particles have, however, so far not been found where expected. The problem has also become more severe by the so-called little hierarchy problem, the fact that on rather general grounds new physics capable of solving the hierarchy problem should have shown up at the LHC, but nothing was observed so far~\cite{Baer:2024ljv}. 

This leads to another way to solve the hierarchy problem, namely mechanisms that naturally produce a tiny value of $\lambda_p$ within the framework of a classically scale-invariant theory. Within QFT one would naturally expect $\lambda_p = {\cal{O}}(1)$ since $H^\dag H$ and $\Phi^\dag \Phi$ are both singlets and since there is no symmetry which protects a tiny value of $\lambda_p$. Things change, however, if one of the scalar fields is not fundamental, but composite. A tree level portal coupling is then absent and an effective portal coupling will be induced by loops involving the fundamental and the composite scalars. This leads at least to some loop suppression of the effective portal term, but can under certain conditions also lead to tiny portal couplings and hierarchies as we will see. 

$\Lambda$CDM is equally successful as the SM, but it has also open issues. One of them is that the underlying theory of gravity, Einstein gravity, is not renormalizable such that quantum effects are not calculable. Another connected question is that cosmic inflation~\cite{Starobinsky:1980te,Sato:1980yn,Linde:1981mu,Linde:1982zj,Albrecht:1982wi} typically rests on some scalar field $X$ with very special parameter choices to allow for ``slow roll'' solutions which match the experimental fact that the equation of state parameter $\omega =  \frac{p}{\rho}$ is close to $ = -1$. Although the idea of inflation is fully consistent with the Planck and BICEP/Keck data of CMB measurements~\cite{Aghanim:2018eyx,Planck:2018jri,BICEP:2021}, on the quantum level one would, however, expect that the scalar inflaton field has a portal term with the SM Higgs field, inducing a hierarchy problem between the electro-weak and Planck scales and endangering the assumed flatness of the inflaton potential. 

We propose a mechanism in this paper which connects the problems of these two very successful theories leading to interesting solutions where the Planck scale emerges dynamically from the breaking of scale invariance, with successful inflation and with a natural explanation of the hierarchy problem. Throughout the paper we will use the SM as our low energy theory, but we would like to emphasize that the underlying mechanisms can easily be generalized to extensions of the SM.

The paper is organized as follows: First, we discuss how the hierarchy problem in a scale invariant extension of the SM
originates and why it differs  from the hierarchy problem when the extension is not based on the classical scale invariance. In section 3 we introduce our Lagrangian composed of the SM in the conformal limit, a hidden sector with conformal symmetry and scale invariant gravity. Here we also discuss the interesting interplay which arises once scale invariant gravity becomes Einstein gravity. In section 4 we discuss the generation of the Planck scale in our model by dimensional transmutation. Furthermore we discuss how a scalar mass for the SM Higgs boson is induced via gravitational interactions. Section 5 discusses how our model leads to successful inflation and how it explains a hierarchy between the electro-weak scale and the Planck scale. Section 6 contains our summary.

\section{The hierarchy problem and scale invariance}

Before presenting our model, we highlight key distinctions concerning the hierarchy problem in scale-invariant theories\footnote{Though scale invariance and conformal invariance are two different invariances, we do not distinguish between two terminologies here most of the time (see for instance~\cite{Nakayama:2013is}).}. Ideally, one would regulate a QFT without breaking any symmetry of the classical Lagrangian. In practice, symmetry-violating regulators are acceptable so long as the correct Ward/Slavnov-Taylor identities are recovered. A regulator that preserves all symmetries exactly would preclude all genuine quantum anomalies and thus cannot exist in general. This includes scale (and conformal) invariance: it is anomalous, with the trace anomaly governed by the beta functions that encode the running of couplings. This observation underlies the common claim~\cite{Callan:1970yg,Symanzik:1970rt} that the scale anomaly necessitates an independent (additive) mass renormalization for scalar fields. This claim is incorrect. Lowenstein and Zimmermann demonstrated already in the 1970s~\cite{Lowenstein:1974qt,Lowenstein:1975rg, Lowenstein:1975rf,Lowenstein:1975ps} in an extended BPHZ framework that the infrared regulator (which they introduce at the beginning) can be  removed to obtain massless theories in perturbation theory. Let us consider in analogy a massless scalar field $\varphi$ coupled with an $U(1)$ gauge field $A_\alpha$:
\al{
{\cal L}_{} =&
-\frac{1}{4} F_{\alpha\beta}^2 +\eta^{\alpha\beta} (D_\alpha \varphi)^\dag D_\beta \varphi
-\lambda_{\varphi}\,(\varphi^\dag\,\varphi)^2\,,
\label{qed}
}
where $D_\alpha=\partial_\alpha+i e A_\alpha$. Classical scale invariance is broken at loop level, as indicated by the trace of the renormalized, improved energy–momentum tensor~\cite{Callan:1970ze,Chanowitz:1972da,Collins:1976yq,Adler:1976zt}. From (\ref{qed}) we get\footnote{The basic tool to rigorously handle   massless theories is the auxiliary mass method (the Lowenstein-Zimmermann equation)~\cite{Lowenstein:1974qt}. Then applying the renormalized action principle of Lowenstein~\cite{Lowenstein:1971jk}, one can derive  (anomalous) Ward identities in massless theories~\cite{Lowenstein:1974qt}. Then  they can be used to obtain  identities for composite operators (normal products). As for the trace anomaly see e.g.~\cite{Clark:1978jx,Piguet:1980nr,Kraus:1991cq}.}
\al{
\Theta^\alpha_{~\alpha} =&
\frac{\beta_e}{2 e}F_{\alpha\beta}^2 -
\beta_{\lambda_{\varphi}}(\varphi^\dag\,\varphi)^2\,,
\label{trace}
}
where $\beta_e$ and $\beta_{\lambda_\varphi}$ are the $\beta$ functions of $e$ and $\lambda_\varphi$; terms vanishing by the equation of motion (EOM) are omitted. The conformal anomaly (\ref{trace}) is exact and scheme-independent. Eq. (\ref{trace}) is consistent with the proof of~\cite{Lowenstein:1974qt,Lowenstein:1975rg, Lowenstein:1975rf,Lowenstein:1975ps}, showing that the conformal anomaly itself does not generate any mass term. It is important to carefully separate unphysical scales introduced by the regulator from physical scales to avoid the results deviating from (\ref{trace}). In cutoff regularization, for example, the quadratic divergence must be canceled by a counterterm chosen so that the renormalized theory obeys (\ref{trace}); this simply constrains the unphysical aspects of the regularization to ensure consistency with the trace anomaly (\ref{trace}) and it does not constitute a physical fine-tuning problem. Interpreting this mass counter term as independent leads to a completely different theory.

We may illustrate the appearance of the second term in (\ref{trace}) in terms of the effective potential {\`a} la  the Coleman-Weinberg  (CW)~\cite{Coleman:1973jx}, which is given at one loop  by
\al{
V_\text{CW}=&
\lambda_{\varphi}(\varphi^\dag\varphi)^2+\frac{1}{2}\beta^{(1)}_{\lambda_{\varphi}}(\varphi^\dag\varphi)^2
\ln\big(\varphi^\dag\varphi/\mu^2\big)
\label{VCW}\,,
}
where $\beta^{(1)}_{\lambda_{\varphi}}$ is the one-loop $\beta$ function, and $\mu$ is a renormalization scale. Using the equation of motion $D_\alpha D^\alpha \varphi =-\partial V_\text{CW}/\partial \varphi$  with its conjugate, we obtain (\ref{trace}) with $\beta_{\lambda_\varphi}$ replaced by $\beta^{(1)}_{\lambda_\varphi}$.
Therefore, the effective potential (\ref{VCW}) is consistent with  the general formula (\ref{trace}). The effective potential (\ref{VCW}) shifts the vacuum expectation value (VEV) away from the origin, $\langle \varphi^\dag\varphi \rangle=v^2/2 \neq 0$, spontaneously breaking scale invariance\footnote{It is important to distinguish shifts of the VEV from non-linear symmetry realization: shifting the VEV gives a non-linear realization with massless orthogonal Goldstone bosons, while the radial (scale) symmetry is anomalous and explicitly broken. Thus, spontaneous scale breaking is compatible with the scale anomaly.}.
Writing $\varphi=(v+h+ i \eta)/\sqrt{2}$ with $\langle h\rangle = \langle \eta\rangle =0$ gives masses $m^2_A=e^2  v^2$ and $m_h^2= \beta^{(1)}_{\lambda_\varphi} v^2/2$. Since (\ref{trace}) is unchanged up to a total derivative absorbable into $\Theta_{\alpha\beta}$, the renormalization of $m^2_A$ and  $m_h^2$ is not an independent one, even though their perturbative loop corrections are divergent\footnote{It is worth noting that the vector boson mass term explicitly appears in (\ref{trace})  in massive QED.}.

The hierarchy problem emerges in scale invariant theories if more than one scalar scale is generated. For illustration we introduce two real scalar fields $ \sigma$ and $s$ to realize the Gildener-Weinberg mechanism~\cite{Gildener:1976ih} which can be achieved by the potential
\al{
V_{\text{GW}}(s,\sigma)=&
\frac{1}{4}\big(\lambda_{\sigma}\sigma^4+\lambda_{s}s^4+ 
\lambda_{s \sigma}\sigma^2 s^2\big)
+\frac{1}{4}\big(\lambda_{s\varphi}\, s^2
+\lambda_{\sigma\varphi}\,\sigma^2\big) \varphi^\dag\varphi\,,
\label{VGW}
}
which is classically scale invariant. As a result, the following term emerges as addition to~(\ref{trace})
\al{
\Delta\Theta^\alpha_{~\alpha} =&
-\frac{1}{4}\big(\beta_{\lambda_{\sigma}}\sigma^4+\beta_{\lambda_{s}}s^4+ 
\beta_{\lambda_{s \sigma}}\sigma^2 s^2\big)
 -\frac{1}{4}\big(\beta_{\lambda_{s\varphi}}\, s^2
+\beta_{\lambda_{\sigma\varphi}}\,\sigma^2\big) \varphi^\dag\varphi\,.
\label{trace2}
}
As before, quantum-induced spontaneous scale-symmetry breaking leaves Eq.~(\ref{trace2}) unchanged. Choose the Gildener–Weinberg parameters such that $\langle s\rangle= 0$ and $\langle \sigma\rangle=v_\sigma \gg v$. The second portal coupling gives then $m_\varphi^2 = \lambda_{\sigma\varphi} v_\sigma^2/4 \gg \beta_{\lambda_\varphi} v^2/2$. Hence, unless $ \lambda_{\sigma\varphi}$ is tiny, $m_\varphi^2 \simeq v_\sigma^2$. The scalar $s$, with mass $O(v_\sigma)$, also contributes to this mass at loop level.
The origin of the physical hierarchy problem in a scale invariant extension of the SM comes therefore from the portal coupling of the Higgs $H$ with other fundamental scalar fields whose scale is much higher than the electro-weak scale. This holds also when fermions are included and their masses are generated by spontaneous scale symmetry breaking. 
We will present a model where a big hierarchy becomes possible by avoiding further fundamental scalars and where two disconnected sub-sectors communicate only via a dynamically generated Planck scale. 

Though studied in flat spacetime, the conformal-anomaly mechanism behind the hierarchy problem is unchanged in curved spacetime\footnote{A possible presence  of a viral current in (\ref{trace}) (see e.g.~\cite{Nakayama:2013is}) does not change the fact of the matter because it is the scale symmetry that dictates the mass renormalization.}. Since the Einstein–Hilbert term is soft in quadratic gravity \cite{Pottel:2020iuz}, massless quadratic gravity should be perturbatively consistent; thus, if the Planck mass $M_\text{Pl}$  is generated spontaneously, its renormalization is not independent.


\section{The model}

Our starting point is a fully scale invariant setting. Therefore we set the SM single mass parameter $\mu_{H}$ to zero such that the SM has no generic scale while the Higgs field $H$ remains a fundamental scalar field. All quadratic divergences, including fermion and gauge boson loops, are absorbed in the renormalization process and do not lead to a hierarchy problem, since there is so far no cutoff or embedding scale. And it will not become a problem if the embedding does not contain any explicit scale. Next we add a non-abelian gauge group $G$ with its gauge-kinetic term $- \frac{1}{2}\Tr F^2$ with a dimensionless gauge coupling and chiral fermions $\psi$. The matter Lagrangian can therefore be written as
\al{ 
{\cal L}_\text{matter} &= {\cal L}_\text{SMGF} + D_\mu H^\dag D^\mu H -\lambda_H(H^\dag H)^2 -\frac{1}{2}\Tr F^2+ \bar{\psi} i \slashed{D}\psi \,, \label{Lmatter} 
}
where ${\cal L}_\text{SMGF}$ stands for the gauge and fermionic parts of the SM, and $F$ is the field-strength tensor of the non-abelian gauge group $G = SU(n_c)$, coupled with the chiral fermions $\psi_i~(i=1,\dots,n_f)$ belonging to the fundamental representation of $SU(n_c)$ and being a SM singlet\footnote{Note that due to the presence of the  fermions the use of the vierbein formalism is silently  understood. But it does not play any role in the following discussions.}. A very simple choice would be $G = SU(3)$ and $n_f = 2$, but other values would be as good.

This overall scale invariant Lagrangian has the interesting and important feature that it has no portal term, since the $G$-sector does not contain any fundamental scalar field. The Lagrangian also does not allow for any other portal by $U(1)$ mixing or a fermionic portal via Yukawa couplings. In other words: There is no portal whatsoever and the SM and $G$-sector constitute completely separated worlds. The Lagrangian (\ref{Lmatter}) depends only on dimensionless couplings: The gauge couplings of the SM and of the $G$-sector, the Higgs self-coupling $\lambda_H$ and Yukawa couplings which are hidden in ${\cal L}_\text{SMGF}$. The $G$-sector very much resembles chiral QCD and its condensation. The running gauge coupling will lead to a gluon condensate $\langle F^2\rangle$ with a scale $\Lambda_F$ and a chiral condensate $\langle\bar{\psi}\psi\rangle$ with the scale $\Lambda_\psi$. 
Here it is important to observe that the chiral condensate $\langle\bar{\psi}\psi\rangle$ breaks both scale invariance and chiral symmetry, while the gluon condensate $\langle F^2\rangle$ only breaks scale invariance. The scale invariance breaking is directly connected to the scale anomaly which is related to the $\beta$-function of the running gauge coupling. 
The contributions of the non-abelian gauge coupling of the $G$-sector are essential since only they ensure that the UV-limit is asymptotically free, reaching the scale invariant limit (fixed point). This shows that the gluon dynamics and the gluon condensate play an essential role in the breaking of scale invariance. But both condensations do not change the fact that the scale invariant SM and the $G$-sector are so far completely disjoint. 

The situation changes in a very interesting way once gravity is included. Having a fully scale invariant setting we actually start from quadratic gravity (QG), which is perturbatively renormalizable~\cite{Stelle:1976gc},
\al{
\frac{{\cal L}_\text{QG} }{\sqrt{-g}}  &= \gamma\,R^2- \kappa\,C_{\mu\nu\alpha\beta}C^{\mu\nu\alpha\beta}\,,
\label{LQG}
}
where $R$ denotes the Ricci curvature scalar, and $C_{\mu\nu\alpha\beta}$ is the Weyl tensor. This adds another two dimensionless parameters, $\gamma$ and $\kappa$. Note that the combination of the matter Lagrangian (\ref{Lmatter}) with the gravity sector (\ref{LQG}) implies an additional renormalizable interaction term $\xi_H H^\dag H R$ between $R$ and the fundamental Higgs $H$ such that the total Lagrangian of our model at the fundamental level reads
\begin{align}
{\cal L}  = {\cal L}_{QG} + {\cal L}_{matter}-\sqrt{-g}\,\xi_H H^\dag H R\,,
\label{Ltotal}
\end{align}
where ${\cal L}_{matter}$  (\ref{Lmatter}) should be made diffeomorphism invariant, accordingly.

The dynamical  scale symmetry breaking in the $G$-sector in a curved spacetime can induce the Planck mass $M_\text{Pl}$ along  with a cosmological constant $\Lambda$ (see e.g. the review~\cite{Adler:1982ri}). In fact the lattice data in a pure Yang-Mills theory are supporting this idea~\cite{Donoghue:2017vvl}. This is also consistent with the important role of the condensate $\langle F^2\rangle$ and of the gauge contributions to the $\beta$-function emphasized above. Restricting the dynamics to the chiral fermions and dynamical chiral symmetry breaking~\cite{Nambu:1960xd,Nambu:1961tp,Nambu:1961fr}, would yield a wrong sign for the Einstein-Hilbert term~\cite{Hill:1991jc,Inagaki:1993ya,Inagaki:1997kz}. We assume therefore that the combined non-perturbative effects, including the important non-abelian gauge contributions of the G sector, produce the correct sign for the Einstein-Hilbert term\footnote{Alternatively one might consider that the chiral fermion $\psi$ is absent.}. The Higgs then $H$ feels the breaking of scale invariance trough gravitational interaction which induces a small Higgs mass term though  the non-minimal coupling (\ref{Ltotal}) at loop levels.

The addition of the  renormalizable gravitational sector (\ref{LQG}) leads therefore in summary to a very interesting interplay between the previously completely isolated sectors in a flat space-time. The total Lagrangian is at the classical level scale invariant and it contains only dimensionless parameters. The condensations in the $G$-sector at a high scale sets via dimensional transmutation the Planck mass.
We will show that this scenario results in successful cosmic inflation and furthermore explains a small  SM Higgs mass, thus explaining the big hierarchy between the electro-weak and Planck scales.

Before we analyze our model in detail we would like to comment on some general aspects of our gravitational sector. First, we would like to point out that the Ricci curvature tensor squared, $R_{\mu\nu\alpha\beta}R^{\mu\nu\alpha\beta}$, is omitted in the Lagrangian (\ref{LQG}), because it (and also $R_{\mu\nu}R^{\mu\nu}$) does not add anything new, as it can be written as a linear combination of $R^2$, $C_{\mu\nu\alpha\beta}C^{\mu\nu\alpha\beta} $ and the Gau\ss-Bonnet term, which is  a surface term. 
A second point concerns the role of Weyl invariance as a generalization of scale invariance in a curved space and potential connections to conformal symmetry~\cite{Nakayama:2013is}. The $\gamma R^2$-term in our total Lagrangian is not Weyl invariant. All other parts of the total Lagrangian are classically Weyl invariant, but develop at the quantum level a Weyl anomaly~\cite{Duff:1993wm}. Weyl invariance is therefore anyway not preserved by our Lagrangian at the quantum level which makes the presence of the $R^2$ term natural. We will not discuss these aspects further and will assume for the rest of this paper that our Lagrangian can be justified as an effective theory emerging from an embedding into some version of conformal gravity. We will show that our effective Lagrangian (\ref{Ltotal}) puts us into the semi-conformal regime, leading to Starobinsky inflation~\cite{Starobinsky:1980te} which works very well, while the quasi-conformal regime studied in~\cite{Salvio:2014soa,Salvio:2019wcp,Salvio:2020axm} would not work in our context. The implicit embedding of our effective Lagrangian (\ref{Ltotal}) into some more general conformal gravity setting implies, however, potential ghost states and we will elaborate on their role in our approach to the hierarchy problem.


\section{Generating the Planck mass and inducing the Higgs mass}

\begin{figure}[ht]
\begin{center}
\includegraphics[width=6.5in]{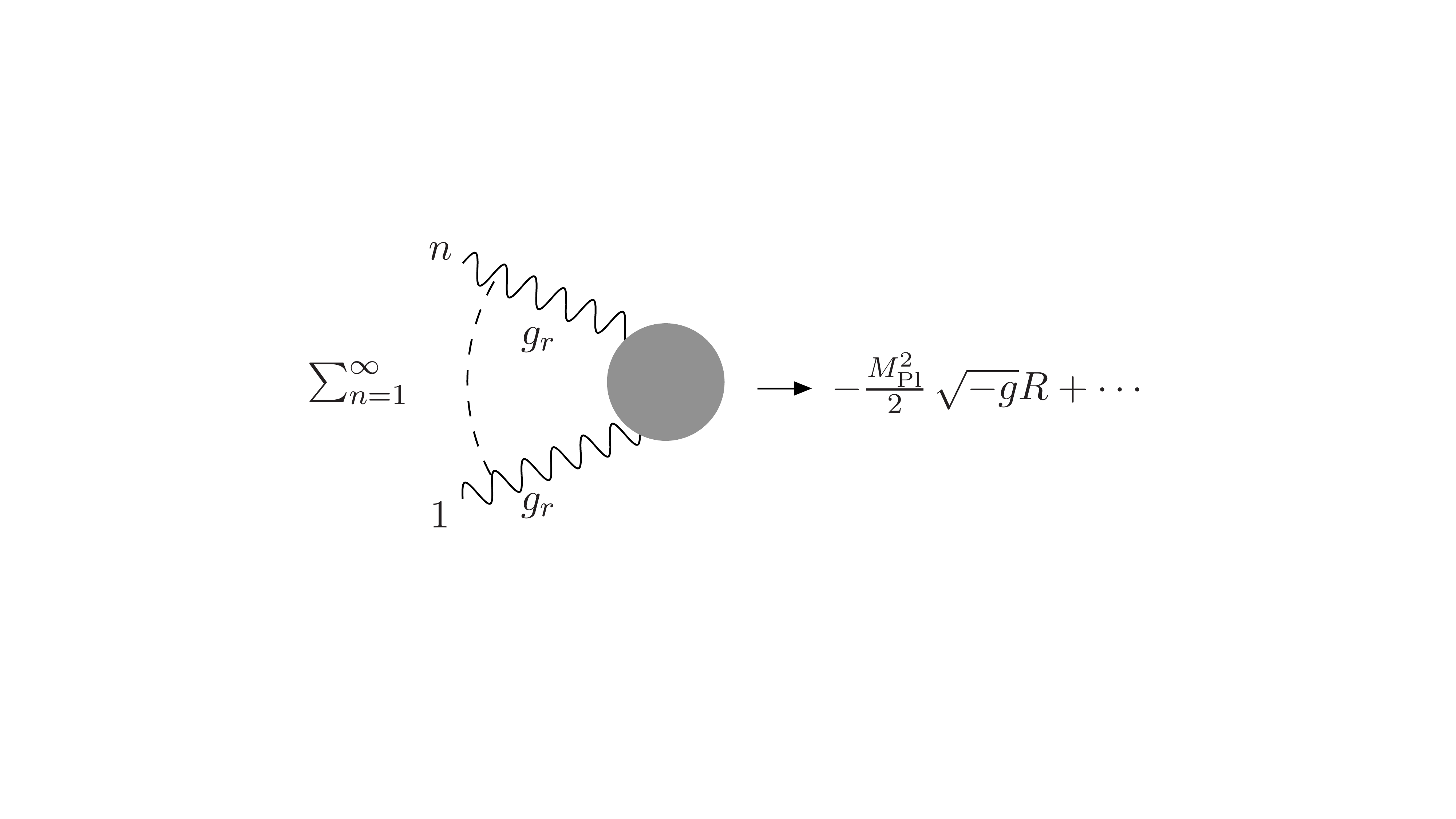}
\vspace{-3cm}
\caption{Generating  the Einstein-Hilbert term, where $g_r$ is  graviton line. The non-perturbative effect is consolidated by the grey disc, and $\cdots$ stand for higher order terms containing more than three derivatives.}
\label{fig-planck}
\end{center}
\end{figure}

The strong dynamics of the $G$-sector breaks scale invariance and produces a robust energy scale.
As stated in the previous section the gravitational interaction undergoes the non-perturbative breaking of scale invariance and as a result the Einstein-Hilbert term is generated~\cite{Adler:1982ri}. This is illustrated in Fig. \ref{fig-planck}, where the non-perturbative effect in the G-sector is consolidated by the grey disc and $g_r$ stands for the graviton lines. As noted, the non-perturbative effect also generates a cosmological constant~\cite{Adler:1982ri}, which is proportional to the VEV of the trace anomaly,  but  we shall postpone this problem and ignore it throughout this paper. (See an interesting discussion of~\cite{Holdom:2007gg} in this regard.)

\begin{figure}[ht]
\begin{center}
\includegraphics[width=3.5in]{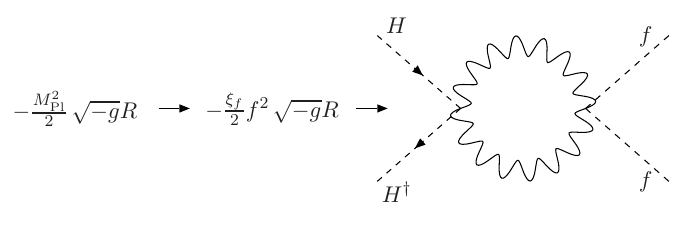}
\caption{Inducing the Higgs mass term. The Einstein-Hilbert term is first formally  replaced by $-\frac{\xi_f}{2}f^2 R$, where $f$ is a Froggatt-Nielsen field. The diagram shows schematically the gravitational linking between $H$ and $f$ at one-loop, which induces  the Higgs mass term once $\xi_f f^2$ is replaced by $M_\text{Pl}^2$.}
\label{fig-non-min}
\end{center}
\end{figure}
The SM Higgs  mass term  originally vanishes due to scale invariance. The dynamical scale  symmetry breaking by the strongly interacting $G$-sector has obviously no direct influence on $H$, but indirectly it induces a mass term via gravitational interactions. We can quantitatively understand the induced generation of the Higgs mass term $\mu_H^2 H^\dag H$ by using the following trick.
We formally  replace the Einstein-Hilbert term $-(M^2_\text{Pl}/2) R$ by $-(\xi_f/2)f^2 R$, where $f$ is a (non-propagating) Froggatt-Nielsen field~\cite{Froggatt:1978nt}. The gravitational interactions in quadratic gravity link the two non-minimal couplings, $-(\xi_f/2)f^2 R$ and  $-\xi_H H^\dag HR$, at loop levels, which is illustrated  in Fig. \ref{fig-non-min} in the case of one-loop. The linking of the two non-minimal couplings induces a portal coupling
\begin{equation}
\frac{{\cal L}_\text{portal}}{\sqrt{-g}} =-\frac{1}{2}\lambda^\text{(ind)}_{f H}
\,f^2 H^\dag H\,.
\label{portal}
\end{equation}

We estimate the size of the portal coupling $\lambda^\text{(ind)}_{f H}$ using the one-loop $\beta$-function~\cite{Salvio:2014soa}
\begin{equation}
\frac{d \lambda_{fH}}{d \ln \mu^2}=\frac{1}{16\pi^2}\ \xi_f\xi_H \Big(\frac{5}{2} f_2^4+\frac{1}{2}f_0^4(6\xi_H+1)(6\xi_f+1)\Big)\,,
\label{beta-portal}
\end{equation} 
where $f_0^2=1/6\gamma$ and $f_2^2=1/2\kappa$, and we have taken into account the fact that any  portal coupling at the fundamental level is absent. When $\xi_f f^2$ is replaced by $M_\text{Pl}^2$, the induced portal coupling (\ref{portal}) becomes the mass term for $H$:
\begin{equation}
\frac{{\cal L}_\text{portal}}{\sqrt{-g}} \to -\mu^2_H H^\dag H 
\end{equation}
with
\begin{equation}
-\mu^2_H \simeq -\frac{\xi_H M^2_\text{Pl}}{256\pi^2} \Big(\frac{5}{\kappa^2} +\frac{1}{9\gamma^2}(6\xi_H+1)(6\xi_f+1)\Big) \times {\cal C}\,,
\label{mus2}
\end{equation}
where we assume that all the non-perturbative effects in the QCD-like sector is consolidated in the parameter $ {\cal C}$, whose absolute value may be in the range  of $O(10^{-1})$ to $O(10)$. The formula (\ref{mus2}) is indeed similar to that of~\cite{Salvio:2014soa}. However, we emphasize that the essential difference is that in our present model, there exits no portal coupling like (\ref{portal}) at the fundamental level, i.e., no local interaction between light and heavy scalar fields; it is induced in a non-perturbative fashion. In this way we can avoid to assume that the portal coupling is of $O(10^{-32})$ if the (fundamental) heavy scalar is a Planck scale field and the light one is the Higgs.

Before we close this subsection let us briefly estimate the effect of the scalaron-$H$ kinetic mixing~\cite{Salvio:2014soa,Kannike:2015apa,Kubo:2022dlx,Alvarez-Luna:2022hka}. Since the scalaron can be very heavy, its kinetic mixing with the Higgs field $H$ may increase the induced $\mu_{H}^2$ given in (\ref{mus2}). We analyse  this effect after the Planck scale and the Higgs mass term have been generated. That is, we include to the original Lagrangian the Einstein-Hilbert term and also the Higgs mass term (\ref{mus2}):
\begin{equation}
\frac{{\cal L}_{\chi H}}{\sqrt{-g}} =-\frac{M^2_\text{Pl}}{2}\,R+\gamma R^2 +g^{\mu\nu}\nabla_\mu H^\dag \nabla_\nu H -\mu_H^2H^\dag H- \xi_H H^\dag HR -\lambda_H(H^\dag H)^2\,,
\label{Lchis}
\end{equation}
where we have suppressed the Weyl tensor squared term because it does not contain the scalaron. A simple way to extract the scalaron degree of freedom in the Lagrangian (\ref{Lchis}) is first to bring $R^2$ term into a linear term:
\begin{equation}
\gamma R^2 \to (M_\text{Pl} /\sqrt{6}) R \,\chi-(m^2_\phi/2)\, \chi^2\,,
\label{aux}
\end{equation}
where $m_\phi^2=M_\text{Pl}^2/ (12 \gamma)$. The new field $\chi$ is an auxiliary field, but propagating and becomes the scalaron in the Jordan frame. To see this, we first define the gravitational fluctuations $h_{\mu\nu}$ around the Minkowski background $\eta_{\mu\nu}$ as $g_{\mu\nu}= \eta_{\mu\nu} +h_{\mu\nu}$, where $h_{\mu\nu}$ describes three  different kinds of (gauge independent) degrees of freedom; massless spin-two, massive spin-two (ghost)~\cite{Stelle:1977ry} and the scalaron \cite{Starobinsky:1980te}. The extraction of $\chi$ from $h_{\mu\nu}$ can be done by introducing traceless and transverse $\hat{h}_{\mu\nu}$ as
\begin{equation}
h_{\mu\nu} =\hat{h}_{\mu\nu}+(\sqrt{2/3})\,\big(\chi/M_\text{Pl}\big)\, \eta_{\mu\nu}\,,
\end{equation}
where $\hat{h}_{\mu\nu}$ describe two different spin-two degrees of freedom. The first term of the Lagrangian (\ref{Lchis}) together with  the first term of (\ref{aux}) gives a canonically normalized kinetic term for $\chi$, while the second term becomes a mass term. To analyze the mixing we write for the complex field $H=(h_1+i h_2)/\sqrt{2}$ where $h_1$ and $h_2$ are real fields and we assume that only $h_1$ acquires  a VEV $v_H$. Then, the quadratic part of ${\cal L}_{\chi H}$ which describes the mixing of $h_1$ with $\chi$ can be written as
\begin{equation}
{\cal L}_{\chi H}^{(\text{mix})} =\frac{1}{2}\big(\partial_\mu \chi \partial^\mu \chi - m_\phi^2  \chi^2\big)+\frac{1}{2}\big(\partial_\mu h_1 \partial^\mu h_1 - m_H^2 h_1^2\big)+\xi_H \Big(\frac{\sqrt{6}v_H}{M_\text{Pl}}\Big)h_1 \,\partial_\mu \partial^\mu \chi\,,
\label{k-mixing}
\end{equation}
where $v_H=\langle h_1\rangle=\sqrt{-\mu_H^2/\lambda}$ and $m_H^2=-2 \mu_H^2$, and it is assumed that $-2 \mu_H^2$ is positive. To transfer the kinetic mixing, i.e., the last term of (\ref{k-mixing}),  into a mass mixing, we first diagonalize the kinetic part and then rescale the fields so that their kinetic terms become canonical. In doing so, we obtain a non-diagonal mass matrix:
\begin{equation}
{\cal M}_{\chi H} = \left(\begin{array}{cc}
\frac{(m_\phi^2+m_H^2)/2}{1-\xi_H \sqrt{6} v_H/M_\text{Pl}}&
\frac{(m_\phi^2-m_H^2)/2}{\big(1-\xi^2_H 6 v_H^2/M_\text{Pl}^2\big)^{1/2}}\\
\frac{(m_\phi^2-m_H^2)/2}{\big(1-\xi^2_H 6 v^2_H/M_\text{Pl}^2\big)^{1/2}}& 
\frac{(m_\phi^2+m_H^2)/2}{1+\xi_H\sqrt{6} v_H/M_\text{Pl}}
\end{array}
\right)\,,
\label{mixing-m}
\end{equation}
with the eigenvalues for $v_H \ll M_\text{Pl}$
\begin{align}
m_{+}^2 & \simeq m_\phi^2+\frac{1}{2}\xi_H^2\, v_H^2/\gamma   +6\, \xi_H^2 m_H^2 v_H^2/M^2_\text{Pl}+ 3 \,\xi_H^4\, v_H^4 / (\gamma M^2_\text{Pl})+O( v_H^6 / M^4_\text{Pl})\,,\\
m_{-}^2 & \simeq m_H^2+O( v_H^6 / M^4_\text{Pl})\,,
\end{align}
implying that the mixing is negligibly small for $v_H \ll M_\text{Pl}$ and $\gamma \sim 10^9$. We therefore will be ignoring the scalaron-Higgs mixing in the following discussions.


\section{Inflation and phenomenological consequences}

If $\gamma R^2$ is present in a theory, inflation works only for  $\gamma\simeq 10^8\,\text{to}\,10^9$ (i.e. $m_\phi\sim 10^{13}\,\text{to}\,10^{14}$ GeV)~\cite{Ema:2017rqn,Pi:2017gih,Salvio:2017xul,Gundhi:2018wyz,Kubo:2018kho,Enckell:2018uic,Kubo:2020fdd,Aoki:2021skm,Cecchini:2024xoq}. Therefore, the scalaron contribution to $\mu_H^2$ (\ref{mus2}) becomes
\begin{equation}
-\mu_H^2 \simeq -\xi_H (6\xi_H+1)(6\xi_\sigma+1)\, \Big(1.6 \times (10^7\,\text{to}\,10^8)\,\text{GeV}\Big)^2\, {\cal C}~\text{for}~\gamma\simeq 10^8\,\text{to}\,10^9\,,
\end{equation}
which is several orders of magnitude larger than the Higgs mass $m_H \simeq 125 $ GeV for $\xi$'s and ${\cal C}$ of $O(1)$. Note, however, that $\mu_H^2$ can be ``naturally'' made small in the semi-conformal regime  in the Higgs sector, i.e., $\xi_H\simeq -1/6$~\cite{Salvio:2014soa,Kannike:2015apa,Salvio:2019wcp,Salvio:2020axm}\footnote{See~\cite{Casarin:2018odz} and references therein for the conformal anomaly of scalar fields.}. Note also that  the non-vanishing term of $\beta_{\xi_H}$ (the one-loop $\beta$ function for $\xi_H$) at $\xi_H=-1/6$ is only the  term $\propto \gamma/\kappa^2$~\cite{Salvio:2014soa}\footnote{In the model we are considering here, the portal coupling is absent, so that the term proportional to it is absent in the $\beta$ function.}. 
As we see from (\ref{mus2}), the spin-two ghost contribution will be $-2\mu_H^2\simeq \Big[(6.2\times 10^{16}/\kappa)\,\text{GeV}\Big]^2 {\cal C}$, implying that $\kappa \simeq 5\times 10^{14}$ to get  $-2 \mu_H^2\simeq m_H^2=(125\,\text{GeV})^2$ (for  ${\cal C}$ of $O(1)$) in the semi-conformal regime. This means, the non-vanishing term of $\beta_{\xi_H}$  at $\xi_H=-1/6$ will be of $O(10^{-19})$, which we may safely neglect, because ${\xi_H}$ practically does not run in the energy range of our interest.
 
At this stage it may be appropriate to clarify the difference between the semi-conformal regime mentioned above and  the quasi-conformal regime considered in~\cite{Salvio:2014soa,Kannike:2015apa,Salvio:2019wcp,Salvio:2020axm}. In  the quasi-conformal regime all the couplings are close to their UV fixed points while $ f_0^2=1/6\gamma \sim \infty$ (accordingly all the non-minimal couplings are $\simeq -1/6$). Therefore, the Starobinsky inflation does  not work in this regime; too small $\gamma$. In contrast to this, all the couplings (except the gauge coupling in the QCD-like sector) in the semi-conformal regime are supposed to be in perturbative regime. Further, we regard our starting renormalizable theory described by (\ref{Ltotal}) as an effective theory below some scale $< \infty$. Though the coupling $f_0^2=1/6 \gamma$ may  grow up to $\infty$ in the infinite energy limit, $\gamma$ and also $\kappa (=1/2 f_2^2)$ vary only slightly in the semi-conformal regime. Using the one-loop $\beta$ functions of~\cite{Salvio:2014soa}, we indeed find
\begin{align}
\delta \left|\frac{\gamma(\mu)}{\gamma(\mu_0)}\right|
&\simeq\frac{1}{16\pi^2} \Big(\frac{5}{2\kappa(\mu_0)}+\frac{5}{36\gamma(\mu_0)}+\frac{5\gamma(\mu_0)}{2\kappa^2(\mu_0)}\Big)|\ln(\mu/\mu_0)| \lsim 4.1\times 10^{-13}\,,\\
\delta \left|\frac{\kappa(\mu)}{\kappa(\mu_0)}\right|
&\simeq  \frac{1}{16\pi^2}\Big( \frac{1081}{120}\Big) \Big(\frac{|\ln(\mu/ \mu_0)|}{\kappa(\mu_0)}\Big)\lsim 1.3\times 10^{-15}
\end{align}
$\text{for}~\mu/\mu_0= 10^{-10}~\text{to}~10^{10}$, where we have used $\gamma(\mu_0)=5\times 10^{8}$ and $\kappa(\mu_0)=10^{15}$. The value of $\gamma(\mu_0)$ is a representative one for the Starobinsky inflation to work, and the value of $\kappa(\mu_0)$ is dictated by the Higgs naturalness. Therefore, we may assume that $\gamma$ and $\kappa$ remain approximately constant in the semi-conformal regime.

\subsection{\texorpdfstring{The coefficient $\gamma$ of the $R^2$-term}{The coefficient gamma of the squared Ricci scalar}}
With the remarks above, we study inflation  in more detail. There are two scalar fields that can actively participate in inflationary dynamics; the scalaron $\chi$ (which we will denote by $\phi$ in the Einstein frame) and the SM Higgs $H$. To proceed, we assume that $\xi_H=-1/6$ for the system to be in the semi-conformal regime:
Undesirable  large $\gamma$ contribution to the induced Higgs mass (\ref{mus2}) is suppressed. The smallness of $\xi_H$ (i.e., $\lesssim O(10)$) means further that the scalaron  can dominate in inflationary dynamics. Consequently, the two-field system for inflation reduces to a single-field system, the Starobinsky inflation~\cite{Starobinsky:1980te,Mukhanov:1981xt,Starobinsky:1983zz}, predicting
\begin{equation}
n_s\simeq1- \frac{2}{N_e}\,,~r\simeq \frac{12}{N_e^2}\,,
\label{spectral}
\end{equation}
where $n_s$,  $r$ and $N_e$ are, respectively, the scalar spectral index, the tensor-to-scalar ratio and the number of e-foldings\footnote{In the presence of the spin-two ghost, the prediction of $r$ will be corrected as we will discuss later on.}.

\begin{figure}[ht]
\begin{center}
\hspace{0.7cm}
\includegraphics[width=3.7in]{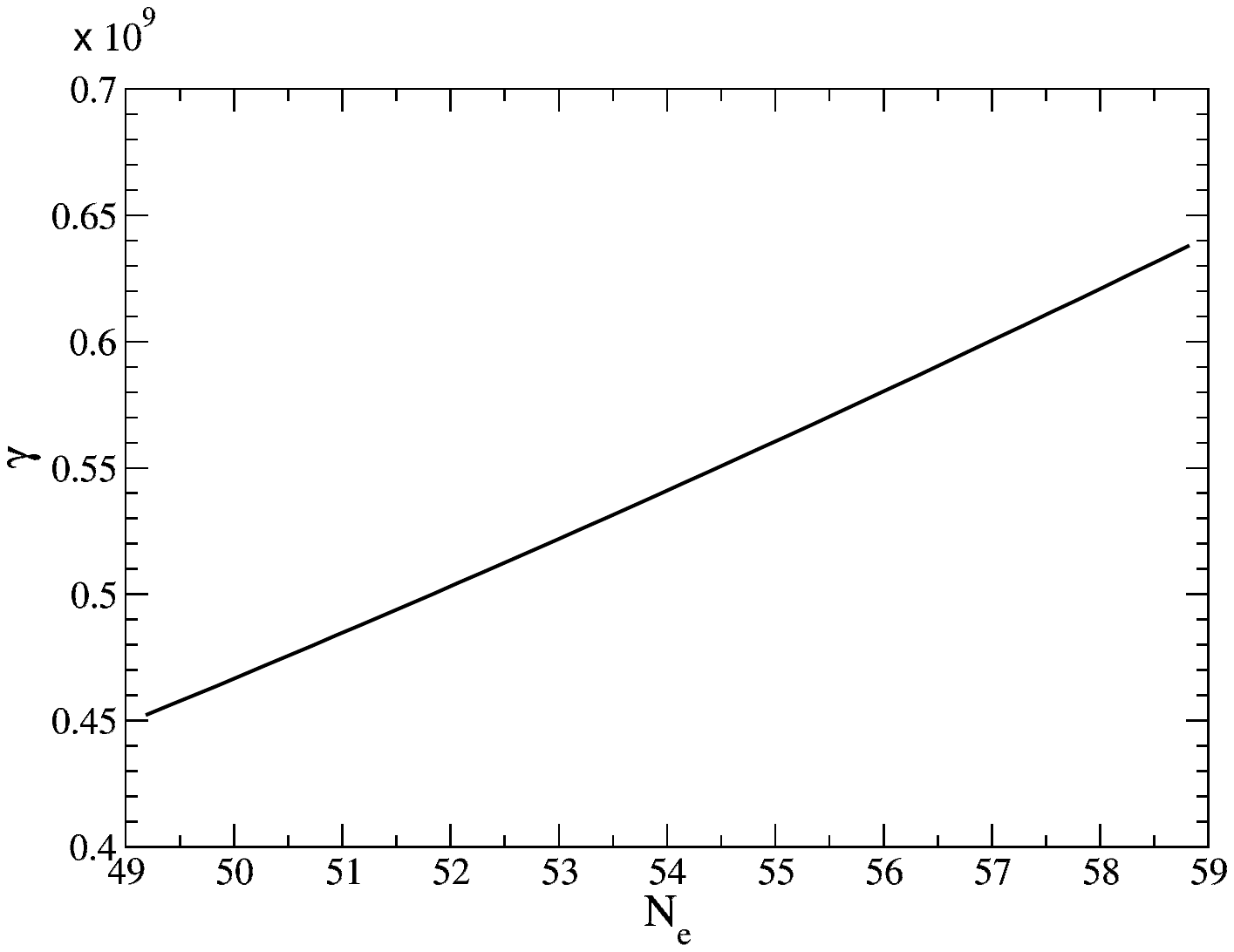}
\caption{ $\gamma$ vs $N_e$.}
\label{gamma1}
\end{center}
\end{figure}

Next we will briefly recall how the constraint on $\gamma$ arises. To this end, we go from the Jordan frame to the Einstein frame, in which the inflationary scalar potential  is given by~\cite{Maeda:1987xf}
\begin{align}
V(\phi) = \frac{M_{\rm Pl}^4}{16\,\gamma}\left(1 - e^{-\sqrt{2/3}\,\phi/M_\text{Pl}}\right)^2\,.
\label{VSphi}
\end{align}
The parameter $\gamma$ enters as an overall factor of the potential, so that  the prediction (\ref{spectral}) does not depend on $\gamma$. The constraint on $\gamma$ comes from the scalar amplitude~\cite{Aghanim:2018eyx,Planck:2018jri}
\begin{equation}
A_s = e^{3.044\pm0.014}\times 10^{-10}\,.
\end{equation}
The amplitude $A_s$ is proportional to $1/\gamma$ because  in the slow-roll approximation it can be written as
\begin{equation}
A_s=\frac{V(\phi_*)}{24\pi^2 \varepsilon_* M^4_\text{Pl}}\,,
\end{equation}
where $\phi_*$ and $\varepsilon_*$ ($\simeq (M^2_\text{Pl}/2) (V'/V)^2$ at $\phi=\phi_*$) are those at the CMB horizon exit~\cite{Aghanim:2018eyx}. Fig. \ref{gamma1} shows the consistent values of $\gamma$ for $N_e \simeq 49\, \text{to}\,59$.

\subsection{\texorpdfstring{The coefficient $\kappa$ of the squared Weyl tensor}{The coefficient kappa of the squared Weyl tensor}}

If the Weyl tensor squared term is present, the inflationary predictions
changes. In particular, the tensor-to-scalar ratio given in (\ref{spectral})~\cite{Deruelle:2012xv,Myung:2014jha,Myung:2015vya,Salvio:2017xul, Ghilencea:2019rqj,Anselmi:2020lpp,Bianchi:2025tyl} and the tensor spectral index $n_t$  get   corrected~\cite{Salvio:2020axm,Anselmi:2020lpp,Kubo:2025jla,Bianchi:2025tyl}:
\begin{equation}
r=16\varepsilon_*\to r= \frac{16\varepsilon_*}{1+2 {\cal H}^2_*/m^2_\text{gh}}\,,~~
n_t = -2 \varepsilon_*  \to n_t=\frac{-2 \varepsilon_* }{1+2 {\cal H}^2_*/m^2_\text{gh}}\,,
\end{equation}
where ${\cal H}_*$ is the Hubble parameter at  horizon exit and $m_\text{gh}=M_\text{Pl}/\sqrt{4\kappa}$ is the mass of the spin-two ghost\footnote{We will  sidestep a contradictory debate about whether an undesirable growth of the scalar part of the ghost perturbation in the superhorizon regime is a gauge artifact~\cite{Deruelle:2010kf,Ivanov:2016hcm,Salvio:2017xul} or not~\cite{DeFelice:2023psw}.}.

The real problem is the fact that the physical  unitarity is violated  in the presence of the spin-two ghost~\cite{Stelle:1977ry}. By the violation of the physical  unitarity  we mean that the probability interpretation of quantum theory fails. This is because the norm of the spin-two ghost states is not positive definite~\cite{Stelle:1977ry}.

The unitarity problem may arise during inflation and also  after inflation. During inflation it is usually assumed  that the ground state of the fluctuations around the background universe (in the Heisenberg picture) is the Bunch-Davies vacuum~\cite{Bunch:1978yq}, which is an empty vacuum state. A crucial  point is that quantum fluctuations of massless (or nearly massless)  modes can become classical field configurations after horizon exit, although their VEV vanishes (see~\cite{Lyth:1984gv,Guth:1985ya,Polarski:1995jg,Lyth:2006qz}). These classical configurations are the seeds of CMB anisotropy and large scale structure of the universe~\cite{Mukhanov:1981xt,Hawking:1982cz,Starobinsky:1982ee,Guth:1982ec,Bardeen:1983qw}. 
As  for the ghost, the negative norm makes it questionable to interpret quantum fluctuations as turning into classical field configurations.  Fortunately, the ghost is massive and therefore its modes will fast die before  horizon exit. In other words, the classicality requirement can not be satisfied~\cite{Starobinsky:1982ee,Lyth:1984gv,Guth:1985ya,Polarski:1995jg,Lyth:2006qz}, meaning that the ghost fluctuations can not become classical, i.e., Wheeler’s ``decoherence without decoherence''~\cite{Polarski:1995jg} can not occur.  As long as the ghost fluctuations are  quantum mechanical virtual excitations, we have  no problem because they do not have any effect on the anisotropy of the universe.

After inflation ends, the universe reheats, and particles are created. This epoch is significant in our discussion because spin-two ghost particles may be produced. There are various interesting ideas to overcome the unitarity problem:

\begin{itemize}
\item First, the pole of the ghost propagator is shifted into the (physical) first sheet of complex four momentum squared and as a result the ghost becomes complex with a pair of conjugate complex masses $m_\text{gh}^\text{c}$ and $(m_\text{gh}^\text{c})^*$ \cite{Lee:1969zze,Lee:1970iw}. Therefore, they may not be produced through collisions among ordinary particles or the decay of ordinary particles, which means that the unitarity problem disappears~\cite{Lee:1969fy,Lee:1969zze,Coleman:1969xz,Lee:1970iw,Cutkosky:1969fq,Anselmi:2017ygm,Anselmi:2018kgz}.
\item Next, even if  the ghost particles can be produced, the problem hinges strongly on whether the ghost is stable or not, or more precisely, whether an asymptotic ghost state exits or not. If there exits no asymptotic ghost state, the theorem of Veltman~\cite{Veltman:1963th} may  be proven~\cite{Grinstein:2008bg,Donoghue:2019fcb,Donoghue:2021eto}, implying that the unstable ghost state does not contribute to the optical theorem, which is a consequence of unitarity.
\item The ghost quantum field might furthermore be "transformed" into a conventional quantum field by introducing a modified inner product in the Hilbert space~\cite{Bender:2007wu,Bender:2008gh,Mannheim:2009zj,Salvio:2015gsi,Salvio:2018crh} (see also~\cite{Kuntz:2024rzu,Salvio:2024joi}). In this case, the ghost particle may be stable and can be produced, without violating the physical unitarity.
\item Finally it should also be noted that we would have  a completely different situation if the ghost were to be confined like the gluon~\cite{Kawasaki:1981gk,Arkani-Hamed:2003pdi,Mukohyama:2009rk}, in which case Eq. (\ref{mus2}) for the induced Higgs mass has to be changed. Here we assume that the ghost particle is fundamental.
\end{itemize}

\noindent Obviously, the constraint on $\kappa$ depends on how  the unitarity problem is overcome by these proposals. For the third option, for instance, the ghost may be a cosmological relic like dark matter, which is subjected to various constraints. There is also a conservative analysis of the ghost problem that leads to a rather stringent viewpoint on $\kappa$:
In~\cite{Kubo:2023lpz,Kubo:2024ysu} the ghost problem has been reanalyzed within the framework of conventional QFT, i.e., respecting in particular the fact that how to integrate the loop momenta in a Feynman diagram (apart from its regularization) is dictated by QFT, leaving no room for arbitrariness.
 
First, using dispersion relations one can derive the K\"all\'en-Lehman  representation of the ghost propagator and show that the asymptotic ghost states with a pair of conjugate complex masses exit~\cite{Kubo:2024ysu}. Consequently, the ghost particles must be stable.
Second, it has been shown in~\cite{Kubo:2023lpz} that the amplitude for the production of the complex  ghost particles through the scattering of ordinary particles does {\it not} always vanish. This is because, in the presence of complex energy, the conventional Dirac delta function that expresses the energy conservation at each vertex of interaction should be generalized to a complex delta function (a complex distribution) which allows such amplitude without violating energy conservation~\cite{Kubo:2023lpz}. The complex delta function defines a sharp threshold $m_\text{thr}=\Re m_\text{gh}^\text{c}-\Im m_\text{gh}^\text{c}$ (for $\Re m_\text{gh}^\text{c} > \Im m_\text{gh}^\text{c}$), below which the ghost production amplitude exactly vanishes. Therefore, $m_\text{thr} > E_\text{max}$ (conservative constraint) is a necessary condition for the ghost to be unable to be produced through scattering with ordinary particles, where $E_\text{max}$ is the maximum kinetic energy in the reheating epoch.
  
To quantify the conservative constraint, we compute below the maximum energy (temperature) $E_\text{max}$ which is assumed to occur just after the end of inflation and is estimated to be $E_\text{max}\simeq\big(\rho_\text{end}^{1/4}T_\text{RH}\big)^{1/2}$\cite{Kolb:1990vq,Chung:1998rq}, where  $T_\text{RH}$ is the reheating temperature and $\rho_\text{end}$ is the  energy density at the end of inflation. Since  $\Re m_\text{gh}^\text{c} \gg \Im m_\text{gh}^\text{c}$ in perturbation theory, we may assume that $m_\text{thr}\simeq m_\text{gh}$. To express the  constraint $m_\text{gh} > E_\text{max}$ quantitatively, we need to know the reheating temperature $T_\text{RH}$. ($\rho_\text{end}$ can be calculated in the slow-roll approximation.) Fortunately, it is possible ~\cite{Liddle:2003as,Martin:2010kz} to constrain the reheating phase and hence $T_\text{RH}$ for a given model without specifying a reheating mechanism. 

\begin{figure}[ht]
\begin{center}
\hspace{0.7cm}
\includegraphics[width=2.8in]{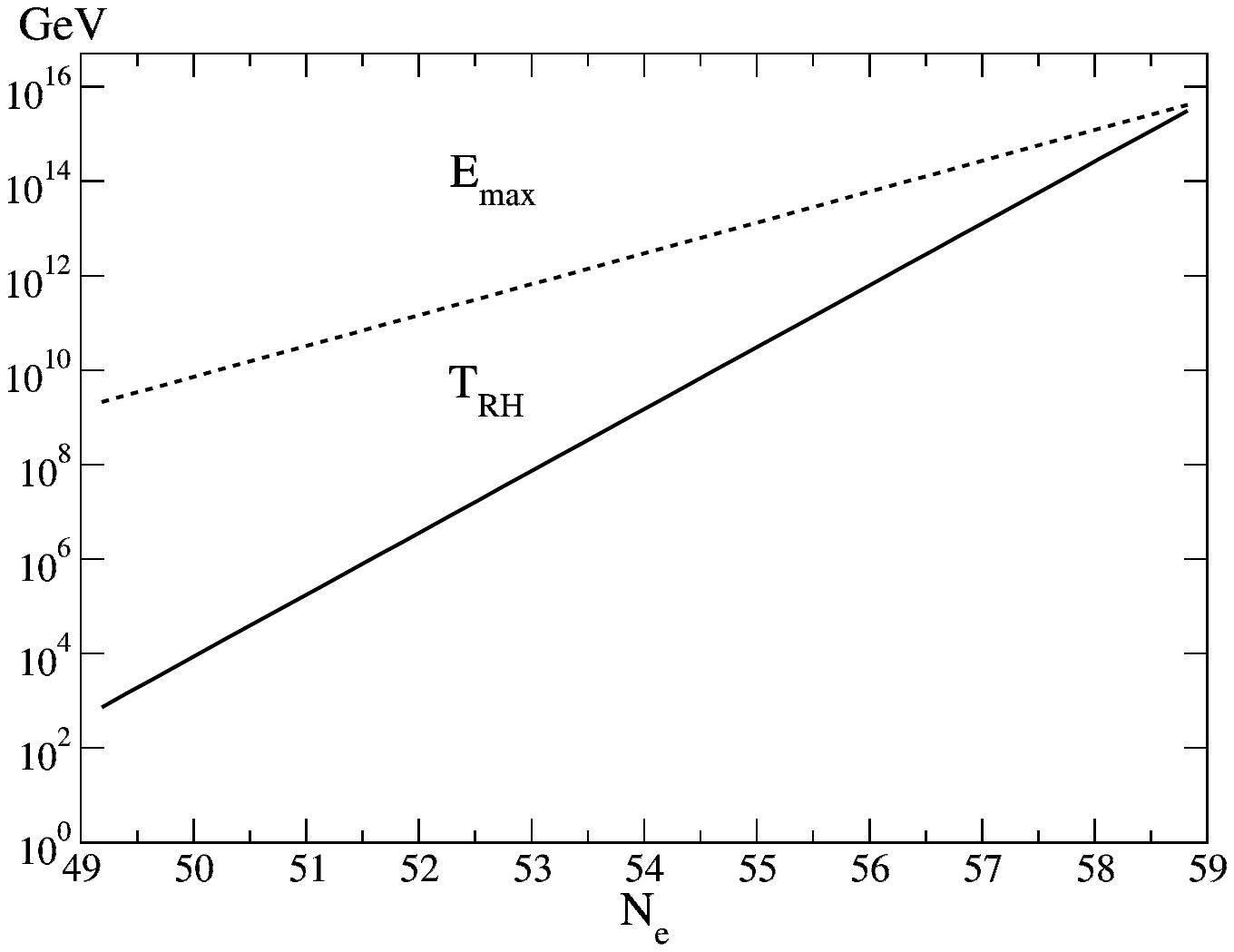}
\includegraphics[width=2.8in]{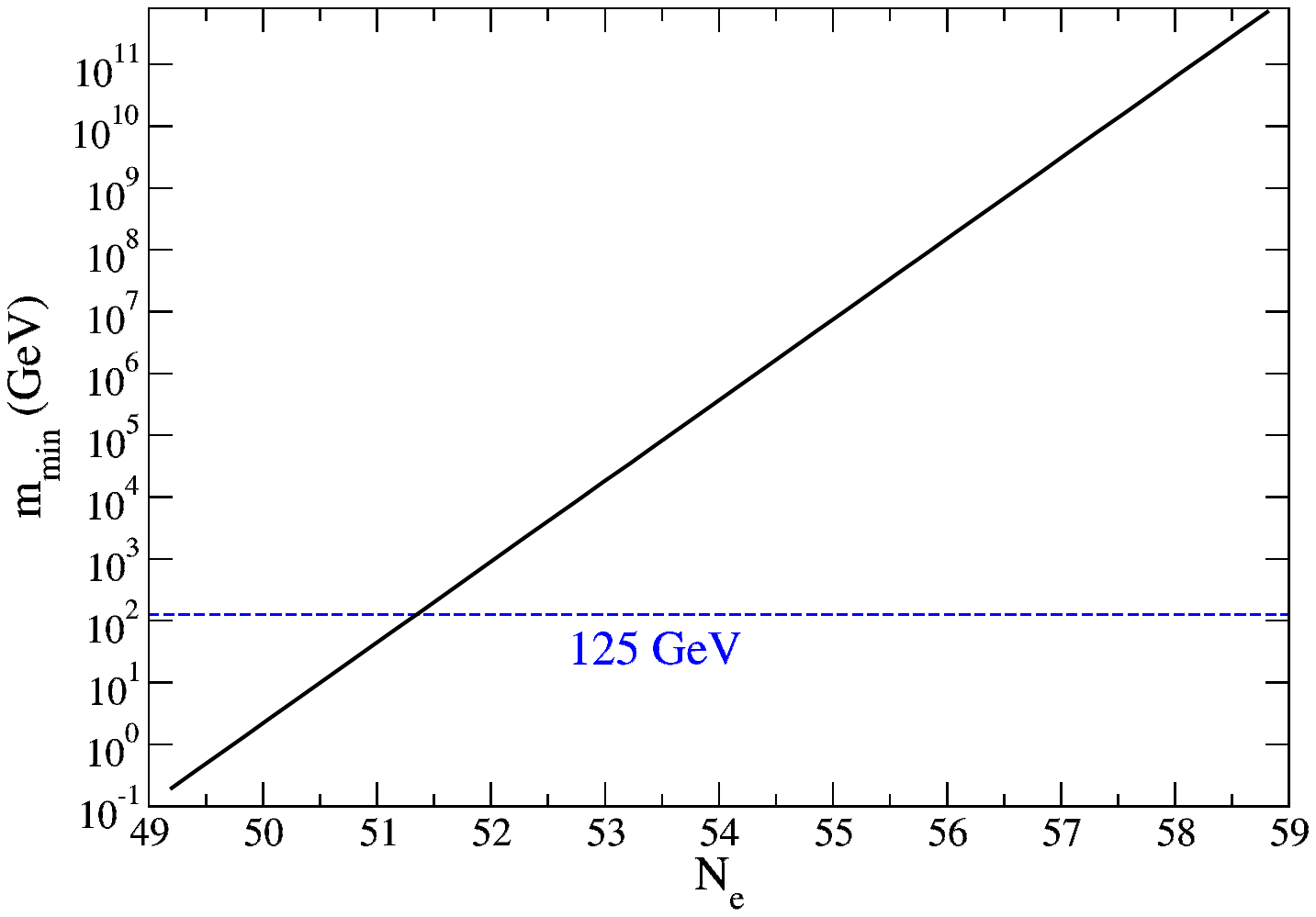}
\caption{$T_\text{RH}$ and $E_\text{max}$ vs $N_e$ (left), and $m_\text{min}$ vs $N_e$ (right)  with ${\cal H}_0= 67.66$ km s$^{-1}$ Mpc$^{-1}$, $k_*=0.002~\text{Mpc}^{-1}$ and $g_\text{RH}=106.75$, where $m_\text{min}$ is the minimum of the induced Higgs mass $(-2 \mu_H^2)^{1/2}$ (for ${\cal C}=1$) and is calculated according to the chain (\ref{chain}). $m_\text{mim}$ does not change when we use ${\cal H}_0= (73.0\pm 1.0)~$ km s$^{-1}$ Mpc$^{-1}$ of \cite{Riess:2021jrx}, because $66.89$ in (\ref{Ne}) should be replaced by $66.81$. The blue line denotes the Higgs mass $125$ GeV.}
\label{TRH1}
\end{center}
\end{figure}
We will follow this idea to find a consistent value of $T_\text{RH}$ for a given $N_e$~\cite{Liddle:2003as,Martin:2010kz,Lozanov:2017hjm,Planck:2018jri}:
\begin{equation}
\begin{split}
N_e &= 66.89-\frac{1}{12}\ln g_\mathrm{RH}+\frac{1}{12}\ln\left(\frac{\rho_\text{RH}}{\rho_\text{end}}\right) +\frac{1}{4} \ln \left(\frac{V(\phi_*)^2}{M_\mathrm{Pl}^4\,\rho_\mathrm{end}}  \right) -\ln \left(\frac{k_*}{a_0 H_0}  \right)\,
\label{Ne}
\end{split}
\end{equation}
with   $\rho_\text{RH}=(\pi^2/30)\, g_\text{RH} \,T_\text{RH}^4$, where $g_\text{RH}$  is the relativistic degrees of freedom at the end of reheating, ${\cal H}_0= (67.66\pm0.42)~$ km s$^{-1}$ Mpc$^{-1}$~\cite{Martin:2010kz,Lozanov:2017hjm,Aghanim:2018eyx}, $a_0=1$, and $k_*=0.002~\text{Mpc}^{-1}$ is the pivot scale set by the Planck mission~\cite{Aghanim:2018eyx,Planck:2018jri}. Further, we notice that the $g_\text{RH}$ dependence cancels in (\ref{Ne}), and using
\begin{equation}
\rho_\text{end}=\frac{V(\phi_\text{end})}{1-\varepsilon_\text{end}/3}= \frac{3V(\phi_\text{end})}{2}~\text{with}~\varepsilon_\text{end}=1\,,
\end{equation}
which can be obtained from the  Friedmann-Lema$\hat{\text{i}}$tre equation and the definition $\varepsilon=-\dot{{\cal H}}/{\cal H}^2$, we finally arrive at
\begin{align}
N_e&\simeq 64.62+\frac{1}{3}\ln \frac{T_\text{RH}}{M_\text{Pl}} +\frac{1}{3} \ln \Big[\frac{2V(\phi_*)}{3V(\phi_\text{end})}\Big] +\frac{1}{6}\ln\Big[ \frac{V(\phi_*)}{M^4_\text{Pl}} \Big]\,.
\label{N2}
\end{align}
Using (\ref{N2}) we can first calculate $T_\text{RH}$ for a given $N_e$, and then $E_\text{max}\simeq\big(\rho_\text{end}^{1/4}T_\text{RH}\big)^{1/2}$, which gives the minimum of $m_\text{gh}$ and hence the maximum of $\kappa$. We then use (\ref{mus2}) to obtain the minimum of the induced Higgs mass $(-2 \mu_H^2)^{1/2}$:
\begin{equation}
\text{Eq. (\ref{N2})}\to T_\text{RH}\to E_\text{max} \to \text{min. of }~m_\text{gh}\to \text{max. of }~\kappa\to \text{min. of }~(-2 \mu_H^2)^{1/2}\,.
\label{chain}
\end{equation}
In Fig. \ref{TRH1} (left) we plot  $T_\text{RH}$ and  $E_\text{max}$ as a function of $N_e$, and in the right panel  the minimum of the induced Higgs mass $(-2 \mu_H^2)^{1/2}$ with ${\cal C}=1$ which we denote by $m_\text{min}$. As we see from the right panel, if ${\cal C}\simeq +1$ and $N_e\lesssim 51.3$, the  electro-weak gauge symmetry breaking can be achieved with the SM Higgs alone without any fine tuning of the Higgs mass. For $N_e\gsim 51.3$ and also for the case with a negative  ${\cal C}$  we need some mechanism to achieve a Higgs naturalness.

The scalaron $\chi$ (inflaton in the Jordan frame), which couples to the trace of the energy momentum tensor, can decay into ghost particles in the reheating epoch: The Weyl anomaly~\cite{Capper:1974ic,Deser:1976yx} induces a coupling $(\chi/M_\text{Pl}) \,C_{\mu\nu\alpha\beta} C^{\mu\nu\alpha\beta} /(4 \pi)^2$. If unitarity violation is only very tiny through the decay, the situation might be tolerated.
We consider an example: $N_e=51.3$ with $\gamma=4.91\times 10^8$ and $\kappa=5.00\times 10^{14}$ which gives $ -2\mu_H^2 =(125~\mbox{GeV})^2,\, \ln( A_s\times 10^{10})=3.044$ and $T_\text{RH}=2.04\times 10^{-13}\, M_\text{PL}$. The (perturbative) partial decay width  into two ghosts for this example is $\sim m_\text{gh}^8/ \big(m_\phi M^6_\text{Pl}(4\pi)^5\big)\simeq 4.3\times 10^{-8}(\gamma^{1/2}/\kappa^4) M_\text{PL}\simeq  1.6 \times 10^{-62}M_\text{PL}$, which should be compared with the Hubble parameter at the end of reheating phase ${\cal H}_\text{RH} = \big(\rho_\text{RH}/(3M^2_\text{Pl})\big)^{1/2} \simeq 1.4\times 10^{-25} M_\text{PL}$.  Therefore, the expansion rate of the universe is  too large for the scalaron to decay into the ghosts. Although the decay process depends on the reheating mechanism, we might expect that the scalaron  decay into spin-two ghosts is negligibly suppressed.

Similarly, even if $m_\text{gh} <  E_\text{max}$, the ghost production rate can be  very small, which may be  tolerated. The ghost production during the reheating phase is a similar process discussed in~\cite{Chung:1998rq}, because  the scalaron  decay into ghosts is negligible. The non-minimal coupling of $H$ in (\ref{Ltotal}) indeed contains $\xi_H\,(m_\text{gh}/M_\text{Pl})^2H^\dag H \varphi_{\mu\nu}\varphi^{\mu\nu}$, which can describe the annihilation of two ghost particles into two Higgs particles, where $\varphi_{\mu\nu}$ is the spin-two ghost field. As it is done in~\cite{Chung:1998rq}, we approximate the thermal average of the annihilation cross section $\langle \sigma |v|\rangle$ to be $ \sim \xi^2_H(m_\text{gh}/M_\text{Pl})^4 /(m^2_\text{gh} 4\pi)$. Then we find that the relic abundance of the ghost can  be estimated as~\cite{Chung:1998rq}
\begin{align}
\Omega_\text{gh} h^2 \sim \frac{\xi^2_H}{4\pi}\Big(\frac{m_\text{gh}}{M_\text{Pl}}\Big)^4 \Big(\frac{T_\text{RH}}{m_\text{gh}}\Big)^7\Big(\frac{106.75}{g_\text{RH}}\Big)^{3/2} \Big(3.3\times 10^{23}\Big) \simeq 9.1\times 10^{-12}\,,
\end{align}
where we have used:
$\xi_H=-1/6\,,~g_\text{RH}=106.75\,,~m_\text{gh}=M_\text{Pl}/\sqrt{ 4 \kappa}= 2.24\times 10^{-8}M_\text{Pl}\, (\ll E_\text{max}=5.89\times 10^{-6}M_\text{Pl})$ and $T_\text{RH}=1.46\times 10^{-8}M_\text{Pl}~ (\mbox{which corresponds to}~N_e=55.0~\mbox{and}~r=2.4\times 10^{-8})$. The value of $m_\text{gh}$ is so chosen, that $(-2\mu_H^2)^{1/2}=125$ GeV. So, the violation of unitarity in this case would be $O(10^{-11})$, which may be tolerated: It is certainly unobservable in the near future.

The smallness of the tensor-to-scalar ratio $r\sim 10^{-8}$ can be traced back to the smallness of the spin-two ghost mass, which is necessary to suppress its contribution to the Higgs mass term. $r$ is so small that it can not be even measured by LiteBIRD~\cite{LiteBIRD:2022cnt} nor by  PICO~\cite{NASAPICO:2019thw}, while our model still predicts Starobinsky inflation for these tiny $r$-values.

\subsection{\texorpdfstring{The Nambu-Goldstone bosons $\pi^a$}{The Nambu-Goldstone bosons}}

The strong dynamics of the QCD-like theory in the G-sector forms  a gauge invariant chiral condensate $\langle \bar{\psi} \psi\rangle$, which breaks the chiral symmetry $SU(n_f)_L\times SU(n_f)_R$ down to  $SU(n_f)_V$, and the associated Nambu-Goldstone (NG) bosons are massless. These massless bosons may have  impacts on cosmology in the early universe.
One can  describe the chiral symmetry breaking in the framework of the Nambu-Jona-Lasinio (NJL) theory~\cite{Nambu:1960xd,Nambu:1961tp,Nambu:1961fr}, which consist of  multi-fermion interactions.
In the self-consistent mean-field approximation of~\cite{Kunihiro:1983ej,Hatsuda:1994pi} (see also~\cite{Holthausen:2013ota}) one  assumes $\langle \bar{\psi} _i\psi_j\rangle \propto \delta_{ij}$ and defines the effective mean field $\sigma$ (order parameter) and the NG boson fields $\pi^a~(a=0,\dots,n_f^2-1)$ as
\begin{align}
\label{varphi}
\sigma \delta_{ij} =- 4 G_{NJL}\,\bar{\psi}_i \psi_j\,, ~~~ 
\pi^a =-2 i G_{NJL}\,\bar{\psi} \gamma_5 \lambda^a\psi\,,
\end{align}
where $G_{NJL}$ is a coupling constant with dimension $-2$ in the NJL theory. By integrating out the fermions in curved space-time with a weakly varying metric one obtains an effective potential   and also  a non-minimal coupling for $\tilde{\sigma}^2=\sigma^2+\pi^a\pi^a$ ~\cite{Inagaki:1993ya,Inagaki:1997kz} (see also~\cite{Larue:2023uyv} for a modern derivation). However, as stated in section 2, the induced non-minimal coupling yields a wrong sign for the Einstein-Hilbert term. Therefore we have to assume that the size of its contribution to the Einstein-Hilbert term is much smaller than that from the gluon condensation.
 
As for the NG bosons, after taking into account the wave function renormalization for $\pi^a$, one finds that the non-minimal coupling for $\pi^a$ is very close to the conformal regime~\cite{Hill:1991jc,Inagaki:1993ya,Inagaki:1997kz}, i.e.,
\begin{align}
{\cal L}_\text{nm} &=
-\frac{\xi_\pi}{2} \pi^a \pi^a\, R
=-\frac{1}{2}\times \Big[\frac{-1}{6}\Big]\Big(1-\epsilon	
\Big)\pi^a \pi^a\, R\,~\mbox{with}~\epsilon=1/\ln \big(
\Lambda_\psi^2/\langle\sigma\rangle^2\big)+\cdots\,,
\label{Lnmm}
\end{align}
where $\Lambda_\psi$ is the scale of the chiral condensate (different from $\langle \sigma \rangle$), and we assume that $\langle \sigma \rangle \ll \Lambda_\psi$ to ensure that $\langle \sigma \rangle \ll M_\text{Pl}$. The small parameter $\epsilon$ is in principle calculable in the framework of the NJL theory.

Since  $\pi^a$ are massless, the scalaron $\chi$ can decay into $\pi$'s kinematically. To study the decay, we need to know the trace of the energy momentum tensor $T_{\mu}^{(\pi)~\mu}$ of $\pi^a$, because $\chi$ couples to  $T_{\mu}^{(\pi)~\mu}$. It  can be found from (\ref{Lnmm}), where the kinetic term for $\pi^a$ is canonically normalized after the wave function renormlaization. We thus arrive at the following conclusion on  $T_{\mu}^{(\pi)~\mu}$:
The dimension two operator  in $T_{\mu}^{(\pi)~\mu}$  is absent because the NG bosons are massless, and those of dimension three and five are also absent  due to an unbroken $SO(n_f^2-1)$ symmetry. The operators of dimension four in  $T_{\mu}^{(\pi)~\mu}$  are multiplied with $\epsilon$, e.g., $\epsilon\, \partial_\mu \pi^a \partial^\mu \pi^a$, while the operators of higher dimensions are suppressed by powers of the form $ \langle\sigma\rangle^p \Lambda_\psi^q $. These operators can be calculated within the framework of the NJL theory. But we leave this calculation to the future project.

As we have seen above, the scalaron decays into the NG bosons $\pi^a$ and
they can be thermalized, but their temperature will be different from that of the SM sector, because their interactions with the SM sector are suppressed by powers of $M_\text{Pl}$ and hence very weak. So we may assume that the temperature of $\pi$, $T_{\pi,\text{RH}}$, at the end of the thermalization phase of the SM sector can be written as
\begin{equation}
T_{\pi,\text{RH}} =\zeta_\pi\,T_\text{RH}\,,
\label{TRH-pi}
\end{equation}
where $T_\text{RH}$ stands for the reheating temperature of the SM sector as before. The constant $\zeta_{\pi}$ is a calculable number in principle, but we leave it unknown here. The thermalized $\pi$'s, which are decoupled from the SM sector, are dark radiation and can contribute to the effective extra relativistic degrees of freedom  $N_\text{eff}$ in the universe~\cite{Ackerman:2008kmp,Steigman:2012ve,Anchordoqui:2012qu}. Under the assumption (\ref{TRH-pi}), applying the conservation of entropy per comoving volume, we can estimate their contribution $\Delta N_\text{eff}$ to $N_\text{eff}$. To this end we have to compute the temperature of $\pi$ at the neutrino decoupling. Using $(n_f^2-1)\, a^3_\text{RH} \,(T_{\pi,\text{RH}})^3 =(n_f^2-1)\, a^3_\nu \,(T_{\pi,\nu})^3$ and $g_{*s}(T_\text{RH})\,a^3_\text{RH } \,T^3_\text{RH} =g_{*s}(T_\nu)\, a^3_\nu \,T^3_\nu$, we first obtain
\begin{equation}
T_{\pi,\nu} =\Big[\frac{g_{*s}(T_\nu)}{g_{*s}(T_\text{RH})}\Big]^{1/3}\, \zeta_\pi\,T_\nu\,.
\end{equation}
Then the $\pi$ contribution to the energy density is
\begin{align}
\rho_\pi 
&=\frac{\pi^2}{30} (n_f^2-1) (T_{\pi,\nu})^4 
= \frac{\pi^2}{30} (n_f^2-1) \Big[\frac{g_{*s}(T_\nu)}{g_{*s}(T_\text{RH})}\Big]^{4/3}\,\zeta_\pi^4\,T_\nu^4\nonumber\\
&=\frac{\pi^2}{30} \Delta N_\text{eff}\Big(\frac{7\times 2}{8}\Big)\, T_\nu^4\,,
\end{align}
which means that
\begin{align}
\Delta N_\text{eff}=\Big[\frac{4(n_f^2-1)}{7}\Big] \Big[\frac{g_{*s}(T_\nu)}{g_{*s}(T_\text{RH})}\Big]^{4/3}\,\zeta_\pi^4 \simeq 0.027\times (n_f^2-1)\zeta_\pi^4\lesssim 0.11\,,
\end{align}
where we have used $g_{*s}(T_\nu)=10.75$ and $g_{*s}(T_\text{RH})=106.75$. The last inequality can be inferred from the Planck constraint $2.99\pm0.17=3.046+\Delta N_\text{eff}$~\cite{Aghanim:2018eyx}, implying that $ (n_f^2-1)\,\zeta_\pi^4\lesssim 4.1$. Therefore, the Planck constraint can be satisfied for $n_f=3$ if e.g. $\zeta_\pi\simeq 0.8$, while $n_f=1$ is a solution for $\zeta_\pi > 1$.


\section{Summary and conclusions}

We study in this paper a potential connection between the generation of the Planck mass by a dynamical breaking of scale invariant gravity and the hierarchy problem of the Standard Model. The hierarchy problem is a problem among explicit scales of scalar operators, which are different by many orders of magnitude~\cite{Peskin:2025lsg}. The electro-weak scale and the Planck mass are vastly different and they relate to completely different physics. One might therefore expect that they are completely independent. A common origin would after all also require to generate one single scale, from which a vastly different other scale would emerge. This seems even more challenging if the first scale is generated dynamically, where the desired hierarchy is also realized dynamically, that is, through interaction (mediation). We showed that gravity mediation in quadratic gravity is an attractive possibility to achieve that, since the basic structure of mediation is fixed by diffeomorphism  invariance and renormalizability. 

We introduced in this paper a model which implements a mechanism where the hierarchy between the Planck and electro-weak scales is by construction a consequence of gravity mediation. We therefore started from the scale invariant Standard Model and added an additional QCD-like $G$-sector which is also scale invariant. The particle content is chosen to be orthogonal such that no scalar, Yukawa or $U(1)$ kinetic mixing portal terms connect the SM and $G$ sectors. A gravitational sector is added as scale invariant, renormalizable quadratic gravity~\cite{Stelle:1976gc}. 

The gauge coupling in the $G$-sector is chosen such that a gluon condensate $\langle F^2\rangle$ with the scale $\Lambda_F$ and a chiral condensate $\langle \bar{\psi}\psi\rangle$ with the scale $\Lambda_\psi$ occur at an energy scale higher than the Planck scale. We have argued that the gluon condensate plays a more essential role in the breaking of scale invariance because it is  the non-abelian nature of the gluon dynamics that dictates ultraviolet as well as infrared physics. Both condensations in a curved spacetime background generate dynamically via dimensional transmutation the Einstein-Hilbert term. Since the chiral condensate yields a wrong sign for the Einstein-Hilbert term, we have assumed that that the total non-perturbative effects in the G sector (dominated by the gluon dynamics) lead to a correct sign for the Einstein-Hilbert term.

These condensations correspond both to a VEV of a local singlet composite operator which can be effectively seen as a composite scalar field. Here it is important to note that tree level portal couplings of the Higgs with these effective scalars  do not exist due to the orthogonality of the fundamental fields in the SM and $G$ sectors. The generation of the Einstein-Hilbert term via dimensional transmutation can be then understood in terms of the non-minimal couplings of the effective scalars, which are generated through gravitational interactions. These non-minimal couplings and that of the fundamental Higgs are successively linked by  gravity at the loop level, inducing a gravitationally suppressed portal term  and consequently a Higgs mass term.
There are two different contributions to the induced Higgs mass  $(-2 \mu^2_H)^{1/2}$; the scalaron and spin-two ghost contributions, each proportional to its mass squared, i.e., $m^2_\phi/M_\text{Pl}$ and  $m^2_\text{gh}/M_\text{Pl}$, respectively. 
 
The size of $m^2_\phi=M^2_\text{Pl}/(12 \gamma)$ is fixed by inflation, because the amplitude of the scalar power spectrum is $\propto 1/\gamma$ and is measured by the Planck mission~\cite{Planck:2018jri} to be $\gamma\sim 10^9$. This indicates at first sight a Higgs mass which is  a few orders of magnitude larger than $125$ GeV. If the Higgs sector is, however, in the semi-conformal regime, i.e. $\xi_H=-1/6$ then this drastically suppresses the large scalaron contribution to $-2 \mu_H^2$. In the semi-conformal regime as opposed to the  quasi-conformal regime~\cite{Salvio:2014soa,Salvio:2019wcp,Salvio:2020axm}, all couplings (except the gauge coupling in the $G$-sector) are in perturbative regime, and importantly, the multi-field system for inflation in our model reduces approximately to a single-field system, the Starobinsky inflation~\cite{Starobinsky:1980te}, however with a very tiny value of $r$ which distinguishes the model from many other models.
 
As for $m^2_\text{gh}=M^2_\text{Pl}/(4 \kappa)$, there is basically no constraint from inflation, but the existence of the spin-two ghost in quadratic gravity causes a serious problem on unitarity~\cite{Stelle:1977ry}.  We mentioned existing ideas to overcome the unitarity problem and subsequently considered ghost production during the reheating phase of the universe in a conservative scenario based on conventional  QFT. We find that we need the spin-two ghost, more precisely its virtual quantum excitation, to get a desired size of the Higgs mass, but its real excitation during the reheating phase of the early universe is so suppressed that the violation of unitarity is extremely tiny at an unobservable small level.
 
The QCD-like $G$-sector might lead to mesons and baryons in the transition from the chiral symmetric phase to the broken phase, which takes place above the Planck scale. The resulting particle density will be diluted during cosmic inflation and should be small. This might, however, lead to a contribution to dark matter, which we will analyze in a future work.

The massless Nambu-Goldstone bosons $\pi^{a}$ associated with the chiral symmetry breaking in the $G$-sector behave as dark radiation and can contribute to the effective extra relativistic degrees of freedom $N_{\text{eff}}$~\cite{Ackerman:2008kmp,Steigman:2012ve,Anchordoqui:2012qu}, yielding possible constraints on $n_f$ in the $G$-sector. Although $\zeta_{\pi}$ (the ratio of the reheating temperature of $\pi^{a}$  to that of the SM particles) is calculable in principle, the computation lies beyond the scope of this paper. Another interesting aspect of $\pi^a$ is that they can be  indeed primordial fluctuations  during inflation.  As they are massless, they can contribute to the non-Gaussianity of the curvature perturbations~\cite{Bartolo:2004if,Maldacena:2002vr}, which  may be sufficiently large that it can be measured in the future~\cite{Achucarro:2022qrl}. Pursuing these computations could lead to very valuable insights into the $\pi^{a}$ phenomenology, but this is left to future work.

We would like to stress that the proposed mechanism is easily generalizable to a wide range of BSM theories with moderate scale differences. Our mechanism should in principle work for theories with TeV-ish scales of new physics, where the scale separation to the electro-weak scale could be understood as a moderate suppression, e.g. via loops. The accommodation of Grand Unified Theories (GUTs) or other models which require other vastly different energy scales requires more consideration. Similarly the cosmological constant problem is beyond the scope of this paper. Both aspects might, however, be investigated in subsequent studies. Finally we also would like to  recall that the sign of the  parameter ${\cal C}$ ($>0$) given in (\ref{mus2}) is crucial for our mechanism  for the gravitationally suppressed Higgs mass to work. Namely, the non-perturbative effect of chiral symmetry breaking is consolidated in ${\cal C}$. We hope that ${\cal C}$ will be available in non-perturbative calculations in the  future.


\section*{Acknowledgments}

This work was supported in part by the MEXT/JSPS KAKENHI Grant Number 23K03383 (J.K.).


\bibliographystyle{JHEP}
\bibliography{biblio.bib}

@article{Piguet:1980nr,
    author = "Piguet, O. and Rouet, A.",
    title = "{Symmetries in Perturbative Quantum Field Theory}",
    reportNumber = "CPT-80-P-1231",
    doi = "10.1016/0370-1573(81)90066-1",
    journal = "Phys. Rept.",
    volume = "76",
    pages = "1",
    year = "1981"
}

@article{Kraus:1991cq,
    author = "Kraus, Elisabeth and Sibold, Klaus",
    title = "{Conformal transformation properties of the energy momentum tensor in four-dimensions}",
    reportNumber = "MPI-PH-91-49",
    doi = "10.1016/0550-3213(92)90314-2",
    journal = "Nucl. Phys. B",
    volume = "372",
    pages = "113--144",
    year = "1992"
}

@article{Clark:1978jx,
    author = "Clark, T. E. and Piguet, O. and Sibold, K.",
    title = "{Supercurrents, Renormalization and Anomalies}",
    reportNumber = "UCB-PTH-78/1",
    doi = "10.1016/0550-3213(78)90064-0",
    journal = "Nucl. Phys. B",
    volume = "143",
    pages = "445--484",
    year = "1978"
}

@article{Lowenstein:1974qt,
    author = "Lowenstein, J. H. and Zimmermann, W.",
    title = "{On the Formulation of Theories with Zero Mass Propagators}",
    reportNumber = "MPI-PAE/PTh 18",
    doi = "10.1016/0550-3213(75)90075-9",
    journal = "Nucl. Phys. B",
    volume = "86",
    pages = "77--103",
    year = "1975"
}

@article{Lowenstein:1971jk,
    author = "Lowenstein, J. H.",
    title = "{Differential vertex operations in Lagrangian field theory}",
    doi = "10.1007/BF01907030",
    journal = "Commun. Math. Phys.",
    volume = "24",
    pages = "1--21",
    year = "1971"
}

@article{Adler:1976zt,
    author = "Adler, Stephen L. and Collins, John C. and Duncan, Anthony",
    title = "{Energy-Momentum-Tensor Trace Anomaly in Spin 1/2 Quantum Electrodynamics}",
    reportNumber = "COO-2220-77-REV, COO-2220-77",
    doi = "10.1103/PhysRevD.15.1712",
    journal = "Phys. Rev. D",
    volume = "15",
    pages = "1712",
    year = "1977"
}

@article{Collins:1976yq,
    author = "Collins, John C. and Duncan, Anthony and Joglekar, Satish D.",
    title = "{Trace and Dilatation Anomalies in Gauge Theories}",
    reportNumber = "COO-2220-88",
    doi = "10.1103/PhysRevD.16.438",
    journal = "Phys. Rev. D",
    volume = "16",
    pages = "438--449",
    year = "1977"
}

@article{Chanowitz:1972da,
    author = "Chanowitz, Michael S. and Ellis, John R.",
    title = "{Canonical Trace Anomalies}",
    reportNumber = "SLAC-PUB-1148",
    doi = "10.1103/PhysRevD.7.2490",
    journal = "Phys. Rev. D",
    volume = "7",
    pages = "2490--2506",
    year = "1973"
}

@article{Callan:1970ze,
    author = "Callan, Jr., Curtis G. and Coleman, Sidney R. and Jackiw, Roman",
    title = "{A New improved energy - momentum tensor}",
    doi = "10.1016/0003-4916(70)90394-5",
    journal = "Annals Phys.",
    volume = "59",
    pages = "42--73",
    year = "1970"
}

@article{Lowenstein:1975ps,
    author = "Lowenstein, J. H.",
    title = "{Convergence Theorems for Renormalized Feynman Integrals with Zero-Mass Propagators}",
    reportNumber = "MPI-PAE/PTh 20/75",
    doi = "10.1007/BF01609353",
    journal = "Commun. Math. Phys.",
    volume = "47",
    pages = "53--68",
    year = "1976"
}

@article{Lowenstein:1975rg,
    author = "Lowenstein, J. H. and Zimmermann, W.",
    title = "{The Power Counting Theorem for Feynman Integrals with Massless Propagators}",
    reportNumber = "MPI-PAE/PTh 6/75",
    doi = "10.1007/BF01609059",
    journal = "Commun. Math. Phys.",
    volume = "44",
    pages = "73--86",
    year = "1975"
}

@article{Froggatt:1978nt,
    author = "Froggatt, C. D. and Nielsen, Holger Bech",
    title = "{Hierarchy of Quark Masses, Cabibbo Angles and CP Violation}",
    reportNumber = "CERN-TH-2519",
    doi = "10.1016/0550-3213(79)90316-X",
    journal = "Nucl. Phys. B",
    volume = "147",
    pages = "277--298",
    year = "1979"
}

@article{Holdom:2007gg,
    author = "Holdom, B.",
    title = "{Massless QCD has vacuum energy?}",
    eprint = "0708.1057",
    archivePrefix = "arXiv",
    primaryClass = "hep-ph",
    doi = "10.1088/1367-2630/10/5/053040",
    journal = "New J. Phys.",
    volume = "10",
    pages = "053040",
    year = "2008"
}

@article{Donoghue:2017vvl,
    author = "Donoghue, John F. and Menezes, Gabriel",
    title = "{Inducing the Einstein action in QCD-like theories}",
    eprint = "1712.04468",
    archivePrefix = "arXiv",
    primaryClass = "hep-ph",
    doi = "10.1103/PhysRevD.97.056022",
    journal = "Phys. Rev. D",
    volume = "97",
    number = "5",
    pages = "056022",
    year = "2018"
}

@article{Adler:1982ri,
    author = "Adler, Stephen L.",
    title = "{Einstein Gravity as a Symmetry-Breaking Effect in Quantum Field Theory}",
    reportNumber = "PRINT-81-0725-REV. (IAS,-PRINCETON), PRINT-81-0725 (IAS,-PRINCETON)",
    doi = "10.1103/RevModPhys.54.729",
    journal = "Rev. Mod. Phys.",
    volume = "54",
    pages = "729",
    year = "1982",
    note = "[Erratum: Rev.Mod.Phys. 55, 837 (1983)]"
}

@article{Hill:1991jc,
    author = "Hill, Christopher T. and Salopek, David S.",
    title = "{Calculable nonminimal coupling of composite scalar bosons to gravity}",
    reportNumber = "FERMILAB-PUB-91-031-T",
    doi = "10.1016/0003-4916(92)90281-P",
    journal = "Annals Phys.",
    volume = "213",
    pages = "21--30",
    year = "1992"
}

@article{ParticleDataGroup:2024cfk,
   author = "Navas, S. and others",
   collaboration = "Particle Data Group",
   title = "{Review of particle physics}",
   doi = "10.1103/PhysRevD.110.030001",
   journal = "Phys. Rev. D",
   volume = "110",
   number = "3",
   pages = "030001",
   year = "2024"
}

@article{Salvio:2024joi,
    author = "Salvio, Alberto",
    title = "{A non-perturbative and background-independent formulation of quadratic gravity}",
    eprint = "2404.08034",
    archivePrefix = "arXiv",
    primaryClass = "hep-th",
    doi = "10.1088/1475-7516/2024/07/092",
    journal = "JCAP",
    volume = "07",
    pages = "092",
    year = "2024"
}

@inproceedings{Baer:2024ljv,
   author = "Baer, Howard",
   title = "{Beyond the Standard Model: An overview}",
   booktitle = "{58th Rencontres de Moriond on QCD and High Energy Interactions}",
   eprint = "2405.00872",
   archivePrefix = "arXiv",
   primaryClass = "hep-ph",
   month = "5",
   year = "2024"
}

@article{Nakayama:2013is,
    author = "Nakayama, Yu",
    title = "{Scale invariance vs conformal invariance}",
    eprint = "1302.0884",
    archivePrefix = "arXiv",
    primaryClass = "hep-th",
    reportNumber = "CALT-68-2910",
    doi = "10.1016/j.physrep.2014.12.003",
    journal = "Phys. Rept.",
    volume = "569",
    pages = "1--93",
    year = "2015"
}

@article{Duff:1993wm,
    author = "Duff, M. J.",
    title = "{Twenty years of the Weyl anomaly}",
    eprint = "hep-th/9308075",
    archivePrefix = "arXiv",
    reportNumber = "CTP-TAMU-06-93",
    doi = "10.1088/0264-9381/11/6/004",
    journal = "Class. Quant. Grav.",
    volume = "11",
    pages = "1387--1404",
    year = "1994"
}

@article{Peskin:2025lsg,
    author = "Peskin, Michael E.",
    title = "{What is the Hierarchy Problem?}",
    eprint = "2505.00694",
    archivePrefix = "arXiv",
    primaryClass = "hep-ph",
    doi = "10.1016/j.nuclphysb.2025.116971",
    journal = "Nucl. Phys. B",
    volume = "1018",
    pages = "116971",
    year = "2025"
}

@article{Kuntz:2024rzu,
    author = "Kuntz, Jeffrey",
    title = "{Unitarity through PT symmetry in quantum quadratic gravity}",
    eprint = "2410.08278",
    archivePrefix = "arXiv",
    primaryClass = "hep-th",
    doi = "10.1088/1361-6382/adf606",
    journal = "Class. Quant. Grav.",
    volume = "42",
    number = "17",
    pages = "175003",
    year = "2025"
}

@article{Bender:2008gh,
    author = "Bender, Carl M. and Mannheim, Philip D.",
    title = "{Exactly solvable PT-symmetric Hamiltonian having no Hermitian counterpart}",
    eprint = "0804.4190",
    archivePrefix = "arXiv",
    primaryClass = "hep-th",
    doi = "10.1103/PhysRevD.78.025022",
    journal = "Phys. Rev. D",
    volume = "78",
    pages = "025022",
    year = "2008"
}

@article{Mannheim:2009zj,
    author = "Mannheim, Philip D.",
    title = "{PT symmetry as a necessary and sufficient condition for unitary time evolution}",
    eprint = "0912.2635",
    archivePrefix = "arXiv",
    primaryClass = "hep-th",
    doi = "10.1098/rsta.2012.0060",
    journal = "Phil. Trans. Roy. Soc. Lond. A",
    volume = "371",
    pages = "20120060",
    year = "2013"
}

@article{Salvio:2015gsi,
    author = "Salvio, Alberto and Strumia, Alessandro",
    title = "{Quantum mechanics of 4-derivative theories}",
    eprint = "1512.01237",
    archivePrefix = "arXiv",
    primaryClass = "hep-th",
    reportNumber = "CERN-PH-TH-2015-290, IFT-UAM-CSIC-15-128",
    doi = "10.1140/epjc/s10052-016-4079-8",
    journal = "Eur. Phys. J. C",
    volume = "76",
    number = "4",
    pages = "227",
    year = "2016"
}

@article{Casarin:2018odz,
    author = "Casarin, Lorenzo and Godazgar, Hadi and Nicolai, Hermann",
    title = "{Conformal Anomaly for Non-Conformal Scalar Fields}",
    eprint = "1809.06681",
    archivePrefix = "arXiv",
    primaryClass = "hep-th",
    doi = "10.1016/j.physletb.2018.10.034",
    journal = "Phys. Lett. B",
    volume = "787",
    pages = "94--99",
    year = "2018"
}

@book{Kolb:1990vq,
    author = "Kolb, Edward W. and Turner, Michael S.",
    title = "{The Early Universe}",
    reportNumber = "FERMILAB-BOOK-1990-01",
    doi = "10.1201/9780429492860",
    isbn = "978-0-429-49286-0, 978-0-201-62674-2",
    publisher = "Taylor and Francis",
    volume = "69",
    month = "5",
    year = "2019"
}

@article{Kannike:2015apa,
    author = {Kannike, Kristjan and H{\"u}tsi, Gert and Pizza, Liberato and Racioppi, Antonio and Raidal, Martti and Salvio, Alberto and Strumia, Alessandro},
    title = "{Dynamically Induced Planck Scale and Inflation}",
    eprint = "1502.01334",
    archivePrefix = "arXiv",
    primaryClass = "astro-ph.CO",
    reportNumber = "IFT-UAM-CSIC-15-015",
    doi = "10.1007/JHEP05(2015)065",
    journal = "JHEP",
    volume = "05",
    pages = "065",
    year = "2015"
}

@article{Alvarez-Luna:2022hka,
    author = "{\'A}lvarez-Luna, Clara and de la Calle-Leal, Sergio and Cembranos, Jos{\'e} A. R. and Sanz-Cillero, Juan Jos{\'e}",
    title = "{Gravitational Coleman-Weinberg mechanism}",
    eprint = "2212.01785",
    archivePrefix = "arXiv",
    primaryClass = "hep-ph",
    doi = "10.1007/JHEP02(2023)232",
    journal = "JHEP",
    volume = "02",
    pages = "232",
    year = "2023"
}

@article{Mukhanov:1981xt,
      author         = "Mukhanov, Viatcheslav F. and Chibisov, G. V.",
      title          = "{Quantum Fluctuations and a Nonsingular Universe}",
      journal        = "JETP Lett.",
      volume         = "33",
      year           = "1981",
      pages          = "532-535",
      note           = "[Pisma Zh.\ Eksp.\ Teor.\ Fiz.\ 33, 549 (1981)]",
      SLACcitation   = "%%CITATION = JTPLA,33,532;%%"
}

@article{Starobinsky:1983zz,
      author         = "Starobinsky, A. A.",
      title          = "{The Perturbation Spectrum Evolving from a Nonsingular
                        Initially De-Sitter Cosmology and the Microwave Background
                        Anisotropy}",
      journal        = "Sov. Astron. Lett.",
      volume         = "9",
      year           = "1983",
      pages          = "302",
      SLACcitation   = "%%CITATION = SALED,9,302;%%"
}

@article{Maeda:1987xf,
    author = "Maeda, Kei-ichi",
    title = "{Inflation as a Transient Attractor in R**2 Cosmology}",
    reportNumber = "UTAP-60-87",
    doi = "10.1103/PhysRevD.37.858",
    journal = "Phys. Rev. D",
    volume = "37",
    pages = "858",
    year = "1988"
}

@article{Holthausen:2013ota,
    author = "Holthausen, Martin and Kubo, Jisuke and Lim, Kher Sham and Lindner, Manfred",
    title = "{Electroweak and Conformal Symmetry Breaking by a Strongly Coupled Hidden Sector}",
    eprint = "1310.4423",
    archivePrefix = "arXiv",
    primaryClass = "hep-ph",
    reportNumber = "KANAZAWA-10-13",
    doi = "10.1007/JHEP12(2013)076",
    journal = "JHEP",
    volume = "12",
    pages = "076",
    year = "2013"
}

@article{Larue:2023uyv,
    author = "Larue, R{\'e}my and Quevillon, J{\'e}r{\'e}mie",
    title = "{The universal one-loop effective action with gravity}",
    eprint = "2303.10203",
    archivePrefix = "arXiv",
    primaryClass = "hep-th",
    reportNumber = "CERN-TH-2023-045",
    doi = "10.1007/JHEP11(2023)045",
    journal = "JHEP",
    volume = "11",
    pages = "045",
    year = "2023"
}

@article{Bianchi:2025tyl,
    author = "Bianchi, Eugenio and Gamonal, Mauricio",
    title = "{Precision predictions of Starobinsky inflation with self-consistent Weyl-squared corrections}",
    eprint = "2506.10081",
    archivePrefix = "arXiv",
    primaryClass = "gr-qc",
    month = "6",
    year = "2025"
}

@article{Kubo:2025jla,
    author = "Kubo, Jisuke and Kuntz, Jeffrey",
    title = "{Primordial gravitational waves in quadratic gravity}",
    eprint = "2502.03543",
    archivePrefix = "arXiv",
    primaryClass = "gr-qc",
    doi = "10.1088/1475-7516/2025/05/093",
    journal = "JCAP",
    volume = "05",
    pages = "093",
    year = "2025"
}

@article{Salvio:2019wcp,
    author = "Salvio, Alberto",
    title = "{Quasi-Conformal Models and the Early Universe}",
    eprint = "1907.00983",
    archivePrefix = "arXiv",
    primaryClass = "hep-ph",
    doi = "10.1140/epjc/s10052-019-7267-5",
    journal = "Eur. Phys. J. C",
    volume = "79",
    number = "9",
    pages = "750",
    year = "2019"
}

@article{Ackerman:2008kmp,
    author = "Ackerman, Lotty and Buckley, Matthew R. and Carroll, Sean M. and Kamionkowski, Marc",
    editor = "Klapdor-Kleingrothaus, Hans Volker and Krivosheina, Irina V.",
    title = "{Dark Matter and Dark Radiation}",
    eprint = "0810.5126",
    archivePrefix = "arXiv",
    primaryClass = "hep-ph",
    doi = "10.1103/PhysRevD.79.023519",
    journal = "Phys. Rev. D",
    volume = "79",
    pages = "023519",
    year = "2009"
}

@article{Anchordoqui:2012qu,
    author = "Anchordoqui, Luis A. and Goldberg, Haim and Steigman, Gary",
    title = "{Right-Handed Neutrinos as the Dark Radiation: Status and Forecasts for the LHC}",
    eprint = "1211.0186",
    archivePrefix = "arXiv",
    primaryClass = "hep-ph",
    doi = "10.1016/j.physletb.2012.12.019",
    journal = "Phys. Lett. B",
    volume = "718",
    pages = "1162--1165",
    year = "2013"
}

@article{Steigman:2012ve,
    author = "Steigman, Gary",
    title = "{Neutrinos And Big Bang Nucleosynthesis}",
    eprint = "1208.0032",
    archivePrefix = "arXiv",
    primaryClass = "hep-ph",
    doi = "10.1155/2012/268321",
    journal = "Adv. High Energy Phys.",
    volume = "2012",
    pages = "268321",
    year = "2012"
}

@article{Riess:2021jrx,
    author = "Riess, Adam G. and others",
    title = "{A Comprehensive Measurement of the Local Value of the Hubble Constant 
    with 1 km/s /Mpc  Uncertainty from the Hubble Space Telescope and the SH0ES Team}",
    eprint = "2112.04510",
    archivePrefix = "arXiv",
    primaryClass = "astro-ph.CO",
    doi = "10.3847/2041-8213/ac5c5b",
    journal = "Astrophys. J. Lett.",
    volume = "934",
    number = "1",
    pages = "L7",
    year = "2022"
}

@article{Ema:2017rqn,
    author = "Ema, Yohei",
    title = "{Higgs Scalaron Mixed Inflation}",
    eprint = "1701.07665",
    archivePrefix = "arXiv",
    primaryClass = "hep-ph",
    reportNumber = "UT-17-04",
    doi = "10.1016/j.physletb.2017.04.060",
    journal = "Phys. Lett. B",
    volume = "770",
    pages = "403--411",
    year = "2017"
}

@article{Pi:2017gih,
    author = "Pi, Shi and Zhang, Ying-li and Huang, Qing-Guo and Sasaki, Misao",
    title = "{Scalaron from $R^2$-gravity as a heavy field}",
    eprint = "1712.09896",
    archivePrefix = "arXiv",
    primaryClass = "astro-ph.CO",
    reportNumber = "YITP-17-135",
    doi = "10.1088/1475-7516/2018/05/042",
    journal = "JCAP",
    volume = "05",
    pages = "042",
    year = "2018"
}

@article{Salvio:2017xul,
    author = "Salvio, Alberto",
    title = "{Inflationary Perturbations in No-Scale Theories}",
    eprint = "1703.08012",
    archivePrefix = "arXiv",
    primaryClass = "astro-ph.CO",
    reportNumber = "CERN-TH-2017-068",
    doi = "10.1140/epjc/s10052-017-4825-6",
    journal = "Eur. Phys. J. C",
    volume = "77",
    number = "4",
    pages = "267",
    year = "2017"
}

@article{Gundhi:2018wyz,
    author = "Gundhi, Anirudh and Steinwachs, Christian F.",
    title = "{Scalaron-Higgs inflation}",
    eprint = "1810.10546",
    archivePrefix = "arXiv",
    primaryClass = "hep-th",
    reportNumber = "FR-PHENO-2018-013",
    doi = "10.1016/j.nuclphysb.2020.114989",
    journal = "Nucl. Phys. B",
    volume = "954",
    pages = "114989",
    year = "2020"
}

@article{Enckell:2018uic,
    author = "Enckell, Vera-Maria and Enqvist, Kari and Rasanen, Syksy and Wahlman, Lumi-Pyry",
    title = "{Higgs-$R^2$ inflation - full slow-roll study at tree-level}",
    eprint = "1812.08754",
    archivePrefix = "arXiv",
    primaryClass = "astro-ph.CO",
    reportNumber = "HIP-2018-37/TH",
    doi = "10.1088/1475-7516/2020/01/041",
    journal = "JCAP",
    volume = "01",
    pages = "041",
    year = "2020"
}

@article{Kubo:2020fdd,
    author = "Kubo, Jisuke and Kuntz, Jeffrey and Lindner, Manfred and Rezacek, Jonas and Saake, Philipp and Trautner, Andreas",
    title = "{Unified emergence of energy scales and cosmic inflation}",
    eprint = "2012.09706",
    archivePrefix = "arXiv",
    primaryClass = "hep-ph",
    doi = "10.1007/JHEP08(2021)016",
    journal = "JHEP",
    volume = "08",
    pages = "016",
    year = "2021"
}

@article{Aoki:2021skm,
    author = "Aoki, Mayumi and Kubo, Jisuke and Yang, Jinbo",
    title = "{Inflation and dark matter after spontaneous Planck scale generation by hidden chiral symmetry breaking}",
    eprint = "2109.04814",
    archivePrefix = "arXiv",
    primaryClass = "hep-ph",
    reportNumber = "KANAZAWA-21-10",
    doi = "10.1088/1475-7516/2022/01/005",
    journal = "JCAP",
    volume = "01",
    number = "01",
    pages = "005",
    year = "2022"
}

@article{Cecchini:2024xoq,
    author = "Cecchini, Chiara and De Angelis, Mariaveronica and Giar{\`e}, William and Rinaldi, Massimiliano and Vagnozzi, Sunny",
    title = "{Testing scale-invariant inflation against cosmological data}",
    eprint = "2403.04316",
    archivePrefix = "arXiv",
    primaryClass = "astro-ph.CO",
    doi = "10.1088/1475-7516/2024/07/058",
    journal = "JCAP",
    volume = "07",
    pages = "058",
    year = "2024"
}

@article{Stelle:1977ry,
    author = "Stelle, K. S.",
    title = "{Classical Gravity with Higher Derivatives}",
    reportNumber = "Print-77-0417 (BRANDEIS)",
    doi = "10.1007/BF00760427",
    journal = "Gen. Rel. Grav.",
    volume = "9",
    pages = "353--371",
    year = "1978"
}

@article{Salvio:2018crh,
    author = "Salvio, Alberto",
    title = "{Quadratic Gravity}",
    eprint = "1804.09944",
    archivePrefix = "arXiv",
    primaryClass = "hep-th",
    reportNumber = "CERN-TH-2018-099",
    doi = "10.3389/fphy.2018.00077",
    journal = "Front. in Phys.",
    volume = "6",
    pages = "77",
    year = "2018"
}

@article{Bartolo:2004if,
    author = "Bartolo, N. and Komatsu, E. and Matarrese, Sabino and Riotto, A.",
    title = "{Non-Gaussianity from inflation: Theory and observations}",
    eprint = "astro-ph/0406398",
    archivePrefix = "arXiv",
    reportNumber = "DFPD-04-A-12",
    doi = "10.1016/j.physrep.2004.08.022",
    journal = "Phys. Rept.",
    volume = "402",
    pages = "103--266",
    year = "2004"
}

@article{Kubo:2018kho,
    author = "Kubo, Jisuke and Lindner, Manfred and Schmitz, Kai and Yamada, Masatoshi",
    title = "{Planck mass and inflation as consequences of dynamically broken scale invariance}",
    eprint = "1811.05950",
    archivePrefix = "arXiv",
    primaryClass = "hep-ph",
    doi = "10.1103/PhysRevD.100.015037",
    journal = "Phys. Rev. D",
    volume = "100",
    number = "1",
    pages = "015037",
    year = "2019"
}

@article{Ghilencea:2019rqj,
    author = "Ghilencea, D. M.",
    title = "{Weyl R$^{2}$ inflation with an emergent Planck scale}",
    eprint = "1906.11572",
    archivePrefix = "arXiv",
    primaryClass = "gr-qc",
    doi = "10.1007/JHEP10(2019)209",
    journal = "JHEP",
    volume = "10",
    pages = "209",
    year = "2019"
}

@article{Lozanov:2017hjm,
    author = "Lozanov, Kaloian D. and Amin, Mustafa A.",
    title = "{Self-resonance after inflation: oscillons, transients and radiation domination}",
    eprint = "1710.06851",
    archivePrefix = "arXiv",
    primaryClass = "astro-ph.CO",
    doi = "10.1103/PhysRevD.97.023533",
    journal = "Phys. Rev. D",
    volume = "97",
    number = "2",
    pages = "023533",
    year = "2018"
}

@article{Martin:2010kz,
    author = "Martin, Jerome and Ringeval, Christophe",
    title = "{First CMB Constraints on the Inflationary Reheating Temperature}",
    eprint = "1004.5525",
    archivePrefix = "arXiv",
    primaryClass = "astro-ph.CO",
    doi = "10.1103/PhysRevD.82.023511",
    journal = "Phys. Rev. D",
    volume = "82",
    pages = "023511",
    year = "2010"
}

@article{Liddle:2003as,
    author = "Liddle, Andrew R and Leach, Samuel M",
    title = "{How long before the end of inflation were observable perturbations produced?}",
    eprint = "astro-ph/0305263",
    archivePrefix = "arXiv",
    doi = "10.1103/PhysRevD.68.103503",
    journal = "Phys. Rev. D",
    volume = "68",
    pages = "103503",
    year = "2003"
}

@article{Chung:1998rq,
    author = "Chung, Daniel J.H. and Kolb, Edward W. and Riotto, Antonio",
    title = "{Production of massive particles during reheating}",
    eprint = "hep-ph/9809453",
    archivePrefix = "arXiv",
    reportNumber = "FERMILAB-PUB-98-217-A, CERN-TH-98-227",
    doi = "10.1103/PhysRevD.60.063504",
    journal = "Phys. Rev. D",
    volume = "60",
    pages = "063504",
    year = "1999"
}

@article{Aghanim:2018eyx,
    author = "Aghanim:2018eyx",
    collaboration = "Planck",
    title = "{Planck 2018 results. VI. Cosmological parameters}",
    eprint = "1807.06209",
    archivePrefix = "arXiv",
    primaryClass = "astro-ph.CO",
    doi = "10.1051/0004-6361/201833910",
    journal = "Astron. Astrophys.",
    volume = "641",
    pages = "A6",
    year = "2020"
}

@article{Linde:1981mu,
      author         = "Linde, Andrei D.",
      title          = "{A New Inflationary Universe Scenario: A Possible
                        Solution of the Horizon, Flatness, Homogeneity, Isotropy
                        and Primordial Monopole Problems}",
      booktitle      = "{QUANTUM COSMOLOGY}",
      journal        = "Phys. Lett.",
      volume         = "108B",
      year           = "1982",
      pages          = "389-393",
      doi            = "10.1016/0370-2693(82)91219-9",
      reportNumber   = "LEBEDEV-81-229",
      SLACcitation   = "%%CITATION = PHLTA,108B,389;%%"
}

@article{Linde:1982zj,
      author         = "Linde, Andrei D.",
      title          = "{Coleman-Weinberg Theory and a New Inflationary Universe
                        Scenario}",
      journal        = "Phys. Lett.",
      volume         = "114B",
      year           = "1982",
      pages          = "431-435",
      doi            = "10.1016/0370-2693(82)90086-7",
      reportNumber   = "LEBEDEV-82-88",
      SLACcitation   = "%%CITATION = PHLTA,114B,431;%%"
}

@article{Albrecht:1982wi,
      author         = "Albrecht, Andreas and Steinhardt, Paul J.",
      title          = "{Cosmology for Grand Unified Theories with Radiatively
                        Induced Symmetry Breaking}",
      journal        = "Phys. Rev. Lett.",
      volume         = "48",
      year           = "1982",
      pages          = "1220-1223",
      doi            = "10.1103/PhysRevLett.48.1220",
      reportNumber   = "UPR-0185T",
      SLACcitation   = "%%CITATION = PRLTA,48,1220;%%"
}

@article{Starobinsky:1980te,
      author         = "Starobinsky, Alexei A.",
      title          = "{A New Type of Isotropic Cosmological Models Without
                        Singularity}",
      journal        = "Phys. Lett.",
      volume         = "B91",
      year           = "1980",
      pages          = "99-102",
      doi            = "10.1016/0370-2693(80)90670-X",
      SLACcitation   = "%%CITATION = PHLTA,B91,99;%%"
}

@article{Coleman:1973jx,
    author = "Coleman, Sidney R. and Weinberg, Erick J.",
    title = "{Radiative Corrections as the Origin of Spontaneous Symmetry Breaking}",
    doi = "10.1103/PhysRevD.7.1888",
    journal = "Phys. Rev. D",
    volume = "7",
    pages = "1888--1910",
    year = "1973"
}

@article{Kubo:2022dlx,
    author = "Kubo, Jisuke and Kuntz, Jeffrey and Rezacek, Jonas and Saake, Philipp",
    title = "{Inflation with massive spin-2 ghosts}",
    eprint = "2207.14329",
    archivePrefix = "arXiv",
    primaryClass = "astro-ph.CO",
    doi = "10.1088/1475-7516/2022/11/049",
    journal = "JCAP",
    volume = "11",
    pages = "049",
    year = "2022"
}

@Article{Nambu:1960xd,
  Title                    = {{Axial vector current conservation in weak interactions}},
  Author                   = {Nambu, Yoichiro},
  Journal                  = {Phys. Rev. Lett.},
  Year                     = {1960},
  Pages                    = {380-382},
  Volume                   = {4},
Doi                      = {10.1103/PhysRevLett.4.380},
  Slaccitation             = {%%CITATION = PRLTA,4,380;%%}
}

@Article{Nambu:1961tp,
  Title                    = {{Dynamical Model of Elementary Particles Based on an
 Analogy with Superconductivity I}},
  Author                   = {Nambu, Yoichiro and Jona-Lasinio, G.},
  Journal                  = {Phys. Rev.},
  Year                     = {1961},
  Pages                    = {345-358},
  Volume                   = {122},

  Doi                      = {10.1103/PhysRev.122.345},
  Slaccitation             = {%%CITATION = PHRVA,122,345;%%}
}

@Article{Nambu:1961fr,
  Title                    = {{Dynamical Model of Elementary Particles Based on an
 Analogy with Superconductivity II}},
  Author                   = {Nambu, Yoichiro and Jona-Lasinio, G.},
  Journal                  = {Phys. Rev.},
  Year                     = {1961},
  Note                     = {[,141(1961)]},
  Pages                    = {246-254},
  Volume                   = {124},
Doi                      = {10.1103/PhysRev.124.246},
  Slaccitation             = {%%CITATION = PHRVA,124,246;%%}
}

@article{Hatsuda:1994pi,
      author         = "Hatsuda, Tetsuo and Kunihiro, Teiji",
      title          = "{QCD phenomenology based on a chiral effective
                        Lagrangian}",
      journal        = "Phys. Rept.",
      volume         = "247",
      year           = "1994",
      pages          = "221-367",
      doi            = "10.1016/0370-1573(94)90022-1",
      eprint         = "hep-ph/9401310",
      archivePrefix  = "arXiv",
      primaryClass   = "hep-ph",
      reportNumber   = "UTHEP-270, RYUTHP-94-1",
      SLACcitation   = "%%CITATION = HEP-PH/9401310;%%"
}

@article{Kunihiro:1983ej,
      author         = "Kunihiro, Teiji and Hatsuda, Tetsuo",
      title          = "{A Selfconsistent Mean Field Approach to the Dynamical
                        Symmetry Breaking: The Effective Potential of the
                        {Nambu-Jona-Lasinio} Model}",
      journal        = "Prog. Theor. Phys.",
      volume         = "71",
      year           = "1984",
      pages          = "1332",
      doi            = "10.1143/PTP.71.1332",
      reportNumber   = "KUNS-705",
      SLACcitation   = "%%CITATION = PTPKA,71,1332;%%"
}

@article{Salvio:2014soa,
    author = "Salvio, Alberto and Strumia, Alessandro",
    title = "{Agravity}",
    eprint = "1403.4226",
    archivePrefix = "arXiv",
    primaryClass = "hep-ph",
    reportNumber = "FTUAM-14-9, IFT-UAM-CSIC-14-021",
    doi = "10.1007/JHEP06(2014)080",
    journal = "JHEP",
    volume = "06",
    pages = "080",
    year = "2014"
}

@article{Inagaki:1993ya,
    author = "Inagaki, T. and Muta, T. and Odintsov, S. D.",
    title = "{Nambu-Jona-Lasinio model in curved space-time}",
    eprint = "hep-th/9306023",
    archivePrefix = "arXiv",
    reportNumber = "HUPD-9314",
    doi = "10.1142/S0217732393001835",
    journal = "Mod. Phys. Lett. A",
    volume = "8",
    pages = "2117--2124",
    year = "1993"
}

@article{Inagaki:1997kz,
    author = "Inagaki, Tomohiro and Muta, Taizo and Odintsov, Sergei D.",
    title = "{Dynamical symmetry breaking in curved space-time: Four fermion interactions}",
    eprint = "hep-th/9711084",
    archivePrefix = "arXiv",
    doi = "10.1143/PTPS.127.93",
    journal = "Prog. Theor. Phys. Suppl.",
    volume = "127",
    pages = "93",
    year = "1997"
}

@article{Bunch:1978yq,
    author = "Bunch, T. S. and Davies, P. C. W.",
    title = "{Quantum Field Theory in de Sitter Space: Renormalization by Point Splitting}",
    doi = "10.1098/rspa.1978.0060",
    journal = "Proc. Roy. Soc. Lond. A",
    volume = "360",
    pages = "117--134",
    year = "1978"
}

@article{Kawasaki:1981gk,
    author = "Kawasaki, Shoichiro and Kimura, Tadahiko",
    title = "{A Possible Mechanism of Ghost Confinement in a Renormalizable Quantum Gravity}",
    reportNumber = "CHIBA-EP-9",
    doi = "10.1143/PTP.65.1767",
    journal = "Prog. Theor. Phys.",
    volume = "65",
    pages = "1767",
    year = "1981"
}

@article{Arkani-Hamed:2003pdi,
    author = "Arkani-Hamed, Nima and Cheng, Hsin-Chia and Luty, Markus A. and Mukohyama, Shinji",
    title = "{Ghost condensation and a consistent infrared modification of gravity}",
    eprint = "hep-th/0312099",
    archivePrefix = "arXiv",
    reportNumber = "HUTP-03-A081, UMD-PPP-04-012",
    doi = "10.1088/1126-6708/2004/05/074",
    journal = "JHEP",
    volume = "05",
    pages = "074",
    year = "2004"
}

@article{Mukohyama:2009rk,
    author = "Mukohyama, Shinji",
    title = "{Ghost condensate and generalized second law}",
    eprint = "0901.3595",
    archivePrefix = "arXiv",
    primaryClass = "hep-th",
    reportNumber = "IPMU-09-0006",
    doi = "10.1088/1126-6708/2009/09/070",
    journal = "JHEP",
    volume = "09",
    pages = "070",
    year = "2009"
}

@article{Hawking:1982cz,
    author = "Hawking, S. W.",
    title = "{The Development of Irregularities in a Single Bubble Inflationary Universe}",
    reportNumber = "Print-83-0015 (CAMBRIDGE)",
    doi = "10.1016/0370-2693(82)90373-2",
    journal = "Phys. Lett. B",
    volume = "115",
    pages = "295",
    year = "1982"
}

@article{Guth:1982ec,
    author = "Guth, Alan H. and Pi, S. Y.",
    title = "{Fluctuations in the New Inflationary Universe}",
    doi = "10.1103/PhysRevLett.49.1110",
    journal = "Phys. Rev. Lett.",
    volume = "49",
    pages = "1110--1113",
    year = "1982"
}

@article{Bardeen:1983qw,
    author = "Bardeen, James M. and Steinhardt, Paul J. and Turner, Michael S.",
    title = "{Spontaneous Creation of Almost Scale - Free Density Perturbations in an Inflationary Universe}",
    reportNumber = "UPR-0202T, EFI-83-13-CHICAGO",
    doi = "10.1103/PhysRevD.28.679",
    journal = "Phys. Rev. D",
    volume = "28",
    pages = "679",
    year = "1983"
}

@article{Donoghue:2019fcb,
    author = "Donoghue, John F. and Menezes, Gabriel",
    title = "{Unitarity, stability and loops of unstable ghosts}",
    eprint = "1908.02416",
    archivePrefix = "arXiv",
    primaryClass = "hep-th",
    reportNumber = "ACFI-T19-08",
    doi = "10.1103/PhysRevD.100.105006",
    journal = "Phys. Rev. D",
    volume = "100",
    number = "10",
    pages = "105006",
    year = "2019"
}

@article{Veltman:1963th,
    author = "Veltman, M. J. G.",
    title = "{Unitarity and causality in a renormalizable field theory with unstable particles}",
    doi = "10.1016/S0031-8914(63)80277-3",
    journal = "Physica",
    volume = "29",
    pages = "186--207",
    year = "1963"
}

@article{Starobinsky:1982ee,
    author = "Starobinsky, Alexei A.",
    title = "{Dynamics of Phase Transition in the New Inflationary Universe Scenario and Generation of Perturbations}",
    doi = "10.1016/0370-2693(82)90541-X",
    journal = "Phys. Lett. B",
    volume = "117",
    pages = "175--178",
    year = "1982"
}

@article{Lyth:1984gv,
    author = "Lyth, D. H.",
    title = "{Large Scale Energy Density Perturbations and Inflation}",
    reportNumber = "Print-84-0373 (LANCASTER)",
    doi = "10.1103/PhysRevD.31.1792",
    journal = "Phys. Rev. D",
    volume = "31",
    pages = "1792--1798",
    year = "1985"
}

@article{Guth:1985ya,
    author = "Guth, Alan H. and Pi, So-Young",
    title = "{The Quantum Mechanics of the Scalar Field in the New Inflationary Universe}",
    reportNumber = "MIT-CTP-1246",
    doi = "10.1103/PhysRevD.32.1899",
    journal = "Phys. Rev. D",
    volume = "32",
    pages = "1899--1920",
    year = "1985"
   }

@article{Polarski:1995jg,
    author = "Polarski, David and Starobinsky, Alexei A.",
    title = "{Semiclassicality and decoherence of cosmological perturbations}",
    eprint = "gr-qc/9504030",
    archivePrefix = "arXiv",
    reportNumber = "LMPM-95-4",
    doi = "10.1088/0264-9381/13/3/006",
    journal = "Class. Quant. Grav.",
    volume = "13",
    pages = "377--392",
    year = "1996"
}

@article{Lyth:2006qz,
    author = "Lyth, David H. and Seery, David",
    title = "{Classicality of the primordial perturbations}",
    eprint = "astro-ph/0607647",
    archivePrefix = "arXiv",
    doi = "10.1016/j.physletb.2008.03.010",
    journal = "Phys. Lett. B",
    volume = "662",
    pages = "309--313",
    year = "2008"
}

@article{NASAPICO:2019thw,
    author = "Hanany, Shaul and others",
    collaboration = "NASA PICO",
    title = "{PICO: Probe of Inflation and Cosmic Origins}",
    eprint = "1902.10541",
    archivePrefix = "arXiv",
    primaryClass = "astro-ph.IM",
    month = "3",
    year = "2019"
}

@article{Lee:1969zze,
    author = {Lee, T. D.},
    title =" {A relativistic complex pole model with indefinite metric,
in Quanta: Essays in Theoretical Physics Dedicated to Gregor Wentzel (Chicago University Press, Chicago, 1970), p. 260, 1969}",
year = "1969"
}

@article{Lee:1969fy,
    author = "Lee, T. D. and Wick, G. C.",
    editor = "Feinberg, G.",
    title = "{Negative Metric and the Unitarity of the S Matrix}",
    doi = "10.1016/0550-3213(69)90098-4",
    journal = "Nucl. Phys. B",
    volume = "9",
    pages = "209--243",
    year = "1969"
}

@article{Lee:1970iw,
author = {Lee, T. D. and Wick, G. C.},
doi = {10.1103/PhysRevD.2.1033},
issn = {05562821},
journal = {Phys. Rev. D},
number = {6},
pages = {1033--1048},
title = {{Finite theory of quantum electrodynamics}},
volume = {2},
year = {1970}
}

@article{Kubo:2023lpz,
    author = "Kubo, Jisuke and Kugo, Taichiro",
    title = "{Unitarity violation in field theories of Lee\textendash{}Wick\textquoteright{}s complex ghost}",
    eprint = "2308.09006",
    archivePrefix = "arXiv",
    primaryClass = "hep-th",
    reportNumber = "YITP-23-104",
    doi = "10.1093/ptep/ptad143",
    journal = "PTEP",
    volume = "2023",
    number = "12",
    pages = "123B02",
    year = "2023"
}

@article{Kubo:2024ysu,
    author = "Kubo, Jisuke and Kugo, Taichiro",
    title = "{Anti-Instability of Complex Ghost}",
    eprint = "2402.15956",
    archivePrefix = "arXiv",
    primaryClass = "hep-th",
    reportNumber = "YITP-24-20",
    doi = "10.1093/ptep/ptae053",
    journal = "PTEP",
    volume = "2024",
    number = "5",
    pages = "053B01",
    year = "2024"
}

@article{Deruelle:2010kf,
    author = "Deruelle, Nathalie and Sasaki, Misao and Sendouda, Yuuiti and Youssef, Ahmed",
    title = "{Inflation with a Weyl term, or ghosts at work}",
    eprint = "1012.5202",
    archivePrefix = "arXiv",
    primaryClass = "gr-qc",
    reportNumber = "YITP-10-105",
    doi = "10.1088/1475-7516/2011/03/040",
    journal = "JCAP",
    volume = "03",
    pages = "040",
    year = "2011"    
}

@article{Ivanov:2016hcm,
    author = "Ivanov, Mikhail M. and Tokareva, Anna A.",
    title = "{Cosmology with a light ghost}",
    eprint = "1610.05330",
    archivePrefix = "arXiv",
    primaryClass = "hep-th",
    reportNumber = "INR-TH-2016-037",
    doi = "10.1088/1475-7516/2016/12/018",
    journal = "JCAP",
    volume = "12",
    pages = "018",
    year = "2016"
}

@article{Anselmi:2018kgz,
    author = "Anselmi, Damiano",
    title = "{Fakeons And Lee-Wick Models}",
    eprint = "1801.00915",
    archivePrefix = "arXiv",
    primaryClass = "hep-th",
    doi = "10.1007/JHEP02(2018)141",
    journal = "JHEP",
    volume = "02",
    pages = "141",
    year = "2018"
}

@article{Callan:1970yg,
    author = "Callan, Jr., Curtis G.",
    title = "{Broken scale invariance in scalar field theory}",
    doi = "10.1103/PhysRevD.2.1541",
    journal = "Phys. Rev. D",
    volume = "2",
    pages = "1541--1547",
    year = "1970"
}

@article{Symanzik:1970rt,
    author = "Symanzik, K.",
    title = "{Small distance behavior in field theory and power counting}",
    reportNumber = "DESY-70-20",
    doi = "10.1007/BF01649434",
    journal = "Commun. Math. Phys.",
    volume = "18",
    pages = "227--246",
    year = "1970"
}

@article{Lowenstein:1975rf,
    author = "Lowenstein, J. H. and Zimmermann, W.",
    title = "{Infrared Convergence of Feynman Integrals for the Massless a**4 Model}",
    reportNumber = "MPI-PAE/PTh 5/75",
    doi = "10.1007/BF01608491",
    journal = "Commun. Math. Phys.",
    volume = "46",
    pages = "105--118",
    year = "1976"
}

@article{Salvio:2020axm,
    author = "Salvio, Alberto",
    title = "{Dimensional Transmutation in Gravity and Cosmology}",
    eprint = "2012.11608",
    archivePrefix = "arXiv",
    primaryClass = "hep-th",
    doi = "10.1142/S0217751X21300064",
    journal = "Int. J. Mod. Phys. A",
    volume = "36",
    number = "08n09",
    pages = "2130006",
    year = "2021"
}

@article{Stelle:1976gc,
    author = "Stelle, K. S.",
    title = "{Renormalization of Higher Derivative Quantum Gravity}",
    reportNumber = "PRINT-76-1059 (BRANDEIS)",
    doi = "10.1103/PhysRevD.16.953",
    journal = "Phys. Rev. D",
    volume = "16",
    pages = "953--969",
    year = "1977"
}

@article{Deruelle:2012xv,
    author = "Deruelle, Nathalie and Sasaki, Misao and Sendouda, Yuuiti and Youssef, Ahmed",
    title = "{Lorentz-violating vs ghost gravitons: the example of Weyl gravity}",
    eprint = "1202.3131",
    archivePrefix = "arXiv",
    primaryClass = "gr-qc",
    reportNumber = "YITP-12-3",
    doi = "10.1007/JHEP09(2012)009",
    journal = "JHEP",
    volume = "09",
    pages = "009",
    year = "2012"
}

@article{Myung:2014jha,
    author = "Myung, Yun Soo and Moon, Taeyoon",
    title = "{Primordial massive gravitational waves from Einstein-Chern-Simons-Weyl gravity}",
    eprint = "1406.4367",
    archivePrefix = "arXiv",
    primaryClass = "gr-qc",
    doi = "10.1088/1475-7516/2014/08/061",
    journal = "JCAP",
    volume = "08",
    pages = "061",
    year = "2014"
}

@article{Myung:2015vya,
    author = "Myung, Yun Soo and Moon, Taeyoon",
    title = "{Scale-invariant tensor spectrum from conformal gravity}",
    eprint = "1501.01749",
    archivePrefix = "arXiv",
    primaryClass = "gr-qc",
    doi = "10.1142/S0217732315501722",
    journal = "Mod. Phys. Lett. A",
    volume = "30",
    number = "32",
    pages = "1550172",
    year = "2015"
}

@article{Anselmi:2020lpp,
    author = "Anselmi, Damiano and Bianchi, Eugenio and Piva, Marco",
    title = "{Predictions of quantum gravity in inflationary cosmology: effects of the Weyl-squared term}",
    eprint = "2005.10293",
    archivePrefix = "arXiv",
    primaryClass = "hep-th",
    doi = "10.1007/JHEP07(2020)211",
    journal = "JHEP",
    volume = "07",
    pages = "211",
    year = "2020"
}

@article{Sato:1980yn,
    author = "Sato, K.",
    title = "{First Order Phase Transition of a Vacuum and Expansion of the Universe}",
    reportNumber = "NORDITA-80-29",
    journal = "Mon. Not. Roy. Astron. Soc.",
    volume = "195",
    pages = "467--479",
    year = "1981"
}

@article{DeFelice:2023psw,
    author = "De Felice, Antonio and Kawaguchi, Ryodai and Mizui, Kotaro and Tsujikawa, Shinji",
    title = "{Starobinsky inflation with a quadratic Weyl tensor}",
    eprint = "2309.01835",
    archivePrefix = "arXiv",
    primaryClass = "gr-qc",
    reportNumber = "YITP-23-106, WUCG-23-09",
    doi = "10.1103/PhysRevD.108.123524",
    journal = "Phys. Rev. D",
    volume = "108",
    number = "12",
    pages = "123524",
    year = "2023"
}

@article{Achucarro:2022qrl,
    author = "Ach\'ucarro, Ana and others",
    title = "{Inflation: Theory and Observations}",
    eprint = "2203.08128",
    archivePrefix = "arXiv",
    primaryClass = "astro-ph.CO",
    month = "3",
    year = "2022"
}

@article{LiteBIRD:2022cnt,
    author = "Allys, E. and others",
    collaboration = "LiteBIRD",
    title = "{Probing Cosmic Inflation with the LiteBIRD Cosmic Microwave Background Polarization Survey}",
    eprint = "2202.02773",
    archivePrefix = "arXiv",
    primaryClass = "astro-ph.IM",
    doi = "10.1093/ptep/ptac150",
    journal = "PTEP",
    volume = "2023",
    number = "4",
    pages = "042F01",
    year = "2023"
}

@article{Maldacena:2002vr,
    author = "Maldacena, Juan Martin",
    title = "{Non-Gaussian features of primordial fluctuations in single field inflationary models}",
    eprint = "astro-ph/0210603",
    archivePrefix = "arXiv",
    doi = "10.1088/1126-6708/2003/05/013",
    journal = "JHEP",
    volume = "05",
    pages = "013",
    year = "2003"
}

@article{BICEP:2021,
    author = "Ade, P. A. R. and others",
    collaboration = "BICEP, Keck",
    title = "{Improved Constraints on Primordial Gravitational Waves using Planck, WMAP, and BICEP/Keck Observations through the 2018 Observing Season}",
    eprint = "2110.00483",
    archivePrefix = "arXiv",
    primaryClass = "astro-ph.CO",
    doi = "10.1103/PhysRevLett.127.151301",
    journal = "Phys. Rev. Lett.",
    volume = "127",
    number = "15",
    pages = "151301",
    year = "2021"
}

@article{Gildener:1976ih,
    author = "Gildener, Eldad and Weinberg, Steven",
    title = "{Symmetry Breaking and Scalar Bosons}",
    reportNumber = "PRINT-76-0068 (HARVARD)",
    doi = "10.1103/PhysRevD.13.3333",
    journal = "Phys. Rev. D",
    volume = "13",
    pages = "3333",
    year = "1976"
}

@article{Planck:2018jri,
    author = "Akrami, Y. and others",
    collaboration = "Planck",
    title = "{Planck 2018 results. X. Constraints on inflation}",
    eprint = "1807.06211",
    archivePrefix = "arXiv",
    primaryClass = "astro-ph.CO",
    doi = "10.1051/0004-6361/201833887",
    journal = "Astron. Astrophys.",
    volume = "641",
    pages = "A10",
    year = "2020"
}

@article{Grinstein:2008bg,
    author = "Grinstein, Benjamin and O'Connell, Donal and Wise, Mark B.",
    title = "{Causality as an emergent macroscopic phenomenon: The Lee-Wick O(N) model}",
    eprint = "0805.2156",
    archivePrefix = "arXiv",
    primaryClass = "hep-th",
    reportNumber = "CALT-68-2684, UCSD-PTH-08-03",
    doi = "10.1103/PhysRevD.79.105019",
    journal = "Phys. Rev. D",
    volume = "79",
    pages = "105019",
    year = "2009"
}

@inproceedings{Coleman:1969xz,
    author = "Coleman, S.",
    title = "{Acausality}",
    booktitle = "{7th International School of Subnuclear Physics (Ettore Majorana): Subnuclear Phenomena}",
    year = "1969"
}

@article{Cutkosky:1969fq,
    author = "Cutkosky, R. E. and Landshoff, P. V. and Olive, David I. and Polkinghorne, J. C.",
    title = "{A non-analytic S matrix}",
    doi = "10.1016/0550-3213(69)90169-2",
    journal = "Nucl. Phys. B",
    volume = "12",
    pages = "281--300",
    year = "1969"
}

@article{Pottel:2020iuz,
    author = "Pottel, Steffen and Sibold, Klaus",
    title = "{On the Perturbative Quantization of Einstein-Hilbert Gravity Embedded in a Higher Derivative Model}",
    eprint = "2012.11450",
    archivePrefix = "arXiv",
    primaryClass = "hep-th",
    doi = "10.1103/PhysRevD.104.086012",
    journal = "Phys. Rev. D",
    volume = "104",
    pages = "086012",
    year = "2021"
}

@article{Capper:1974ic,
    author = "Capper, D. M. and Duff, M. J.",
    title = "{Trace anomalies in dimensional regularization}",
    doi = "10.1007/BF02748300",
    journal = "Nuovo Cim. A",
    volume = "23",
    pages = "173--183",
    year = "1974"
}

@article{Deser:1976yx,
    author = "Deser, Stanley and Duff, M. J. and Isham, C. J.",
    title = "{Nonlocal Conformal Anomalies}",
    reportNumber = "Print-76-0374 (KING S COLL.)",
    doi = "10.1016/0550-3213(76)90480-6",
    journal = "Nucl. Phys. B",
    volume = "111",
    pages = "45--55",
    year = "1976"
}

@article{Donoghue:2021eto,
    author = "Donoghue, John F. and Menezes, Gabriel",
    title = "{Ostrogradsky instability can be overcome by quantum physics}",
    eprint = "2105.00898",
    archivePrefix = "arXiv",
    primaryClass = "hep-th",
    reportNumber = "ACFI-T21-05",
    doi = "10.1103/PhysRevD.104.045010",
    journal = "Phys. Rev. D",
    volume = "104",
    number = "4",
    pages = "045010",
    year = "2021"
}

@article{Bender:2007wu,
    author = "Bender, Carl M. and Mannheim, Philip D.",
    title = "{No-ghost theorem for the fourth-order derivative Pais-Uhlenbeck oscillator model}",
    eprint = "0706.0207",
    archivePrefix = "arXiv",
    primaryClass = "hep-th",
    reportNumber = "PREPRINT-LA-UR-07-3525",
    doi = "10.1103/PhysRevLett.100.110402",
    journal = "Phys. Rev. Lett.",
    volume = "100",
    pages = "110402",
    year = "2008"
}

@article{Anselmi:2017ygm,
    author = "Anselmi, Damiano",
    title = "{On the quantum field theory of the gravitational interactions}",
    eprint = "1704.07728",
    archivePrefix = "arXiv",
    primaryClass = "hep-th",
    doi = "10.1007/JHEP06(2017)086",
    journal = "JHEP",
    volume = "06",
    pages = "086",
    year = "2017"
}

\end{document}